\documentclass[12pt]{article}
\usepackage[left=2.5cm, right=2.5cm, top=3cm]{geometry}
\usepackage{bm}
\usepackage{color}

\usepackage{amsmath}
\usepackage{amssymb}
\usepackage{amsthm}
\usepackage[linesnumbered,ruled,vlined]{algorithm2e}
\usepackage{subcaption}
\usepackage{float}
\usepackage{booktabs}
\usepackage{multirow}
\usepackage{mathtools}

\usepackage{bbm}
\usepackage[hidelinks,colorlinks=true,linkcolor=blue,citecolor=blue]{hyperref}

\usepackage{graphicx}
\usepackage{subcaption}
\usepackage{natbib}
\setcitestyle{authoryear,open={(},close={)}}
\usepackage{booktabs}
\usepackage{amsfonts}
\usepackage{amssymb}
\usepackage{dsfont}
\usepackage{eucal}
\usepackage{multirow}
\usepackage[dvipsnames]{xcolor}
\usepackage{caption}
\usepackage{setspace}

\theoremstyle{plain}

\newtheorem{proposition}{Proposition}
\newcommand{\bcolch}{\color{black}}

\newcommand{\ech}{\color{black}}
\usepackage[most]{tcolorbox}
\theoremstyle{definition}

\DeclareMathOperator*{\argmax}{arg\,max}

\newcommand{\ind}{\mbox{$\perp\!\!\!\perp \,$}}
\newcommand{\nind}{\mbox{$\perp\!\!\!\perp\!\!\!\!\!\! \setminus \;$}}

\newcommand{\E}{\mathbb{E}}
\newcommand{\Pbb}{\mathbb{P}}

\title{Bayesian Controlled FDR Variable Selection via Parameter-Expanded Latent Knockoffs} \author{Lorenzo Focardi-Olmi$\dag$\footnote{Department of Statistics, Rice University, Houston, TX, USA}, Anna Gottard\footnote{Department of Statistics, Computer Science, Application ``G.Parenti",
University of Florence, Italy}, Michele Guindani\footnote{Department of Biostatistics,
University of California at Los Angeles, CA, USA}\\ and\\ Marina Vannucci$^*$}
\date{}

\begin{document}

\maketitle

\begin{abstract}
In many research fields, researchers aim to identify significant associations between a set of explanatory variables and a response while controlling the false discovery rate (FDR). The Knockoff filter has been recently proposed in the frequentist paradigm to introduce controlled noise in a model by cleverly constructing copies of the predictors as auxiliary variables. In this paper, we develop a fully Bayesian generalization of the classical model-X knockoff filter for normally distributed covariates.  {In our approach, we consider a joint model for the covariates and the response, where the conditional independence structure of the covariates is captured through a Gaussian graphical model and used to define a latent knockoff layer through a parameter-expanded representation of the response model. Estimating the covariate graph informs the knockoff construction and improves inference on the covariate effects.} We use a modified spike-and-slab prior on the regression coefficients, which avoids the increase of the model dimension typical of the classical knockoff filter. {We also address extensions to settings with non-Gaussian responses.} Our model performs variable selection using an upper bound on the posterior probability of non-inclusion. {We show that the induced latent knockoff layer defines valid Gaussian model-X knockoffs under the proposed construction and that the resulting procedure controls the Bayesian FDR at an arbitrary level, in finite samples, if the distribution of the covariates is fully known; under an estimated graphical structure, it satisfies an asymptotic FDR guarantee.} We use simulated data to demonstrate that our proposal increases the stability of the selection with respect to classical knockoff methods, as it relies on the entire posterior distribution of the latent knockoff variables instead of a single sample. With respect to Bayesian variable selection methods, we show that our selection procedure achieves comparable or better performances, while maintaining control over the FDR. Finally, we show the usefulness of the proposed model with an application to real data.\\
\end{abstract}


\section{Introduction}\label{Sec:1}
In recent years, variable selection has emerged as a crucial task at the forefront of statistics and machine learning, since it enables searching for associations between an outcome variable and a large amount of possible explanatory variables, when only a few of them are truly relevant. Hence, there is a need for methods that can accurately identify the subset of variables that affect the outcome. Classical approaches for variable selection typically rely on constrained optimization of the likelihood function \citep[see, among others,][]{tibshirani1996lasso, fan2001scad, zou2005elnet}. In the Bayesian approach, variable selection can be performed by employing several classes of prior distributions \citep[for a review, see]{vannucci2021}. For regression models, spike-and-slab priors, proposed for the first time by \cite{mitchell1988ssl}, impose mixture distributions on the regression coefficients,  with latent binary indicators that identify the relevant (active) predictors \citep{george1997bvs, brownetal98}. Another approach involves the use of shrinkage priors to adaptively shrink coefficients towards zero \citep{park2008blasso, Carvalho2010}. 

A fundamental challenge in variable selection is controlling the False Discovery Rate (FDR) to ensure the reliability and replicability of research findings. False discoveries refer to situations where a statistical procedure incorrectly identifies a variable as relevant when, in reality, its effect is merely due to random chance or noise in the data. Several multiple testing procedures have been proposed  to control  the FDR - that is, the expected proportion of inactive variables mistakenly identified as active, among all those selected as active \citep[see, for instance,][]{benjamini1995fdr, benjamini2001yek, benjamini2005mult}. Those procedures typically rely on computing $p$-values, which is not always an easy task \citep{candes2018gold}. Alternative methodologies have been developed by \cite{xiang2021gm}, who propose a method that leverages a pair of mirror variables for each predictor to provide asymptotic control on the FDR, and by \cite{chenguang2022split}, who employ a data-splitting procedure to accomplish the same goal with minimal computational effort. 

In Bayesian settings, the usual definition of FDR cannot be used, since both the set of active and inactive variables are considered random. Instead, the Bayesian FDR (BFDR) is defined as an expected FDR conditionally upon the observed data, i.e., computed with respect to the posterior distribution of the sets of active/inactive variables given the observed data \citep{muller2006fdr, whittemore2007bfdr}.  An advantage of Bayesian methods is that they can facilitate multiplicity adjustments through the choice of the prior. For example, in the context of variable selection, \cite{scott2010bayes} show that a spike-and-slab prior with a Beta-Bernoulli prior on the selection indicator achieves an asymptotic multiplicity control. \cite{muller2006fdr} and \cite{guindani2009bayesian} prove control of the Bayesian FDR by characterizing the FDR problem as a decision problem where the optimal decision rule is based on thresholding the marginal posterior probabilities. More recently, \cite{castillo2020spike} prove control of the Bayes FDR for general empirical Bayes multiple testing procedures. In practical settings, the optimal threshold for controlling the FDR corresponds to the minimum marginal probability of inclusion that ensures that the BFDR is less than a  desired level \citep{newton2004bfdr}. Therefore, this procedure relies on the estimates of these probabilities, which are typically obtained from the MCMC output.

Recent classical techniques to control the FDR do so by introducing carefully designed perturbations to the data. For example, \cite{barber2015knockoff} propose the \textit{knockoff filter}, a powerful method that achieves finite sample FDR control in fixed-design, low-dimensional linear models.  The key idea is to add controlled noise in the form of cleverly constructed copies of the predictors. These copies,  called \textit{knockoffs}, are constructed to mimic the correlation structure of the true covariates while being statistically independent of the response. A linear model is then fitted to this augmented design, and variable selection is performed using a function of the coefficients of both the original and the knockoff variables. Essentially, knockoff variables act as negative controls. 
This framework has been extended to high-dimensional fixed-design settings in \citet{barber2019knockoff}, and is often referred to as the \textit{fixed-X knockoff} procedure.  The \textit{model-X knockoff filter} of \citet{candes2018gold} further extends the approach, by avoiding assumptions on the relation between the response and the covariates. Instead, it requires full knowledge of the joint distribution of the covariates. Alternative perspectives have been investigated by \cite{sesia2018hmm}, who provide an efficient exact sampling of the knockoff variables when the covariates are distributed as a hidden Markov model, and by \cite{bates2021metr}, who apply a modified version of the algorithm proposed by \cite{candes2018gold} to sample knockoffs exactly in the case of known independence structure of the covariates. More recently, \cite{berti2023new} provide a new characterization of the joint distribution of the observed covariates and the knockoffs, which they define in terms of copulas. When the covariates follow a joint Gaussian distribution, their approach recovers the same solution as in \cite{candes2018gold}.

One of the main limitations of the classical model-X knockoff filter is that it depends on the quality of a single random draw of knockoff variables: different draws, from the same set of predictors, may yield different selected sets. To reduce this variability, \citet{ren2021derandomizing} combine the knockoff filter with the stability-selection framework of \citet{meinshausen2010stability}, yielding a more stable selection of active variables.

In this work, we take a fully Bayesian approach that generalizes the model-X knockoff filter and builds stability directly into the inference process. {Rather than conditioning on a single realized knockoff matrix, we introduce a latent knockoff layer through a parameter-expanded representation of a baseline response model and propagate uncertainty in that layer through posterior inference.} We further assume a Gaussian graphical model to model the covariates' dependence structure and impose a continuous spike-and-slab prior on the precision matrix \citep{wang2015graph}. Estimating a graphical model on the covariates allows us to obtain a coherent model that does not require perfect knowledge of the precision matrix. Moreover, unlike plug-in estimators, our Bayesian approach can naturally incorporate prior information about the conditional dependence structure of the covariates into the model. As argued by \citet{barber2020rob}, this leads to higher-quality knockoffs. For the regression coefficients,  we adopt the mixture spike-and-slab prior suggested by \citet{candes2018gold}, which avoids the increase of the model dimension typical of the classical knockoff filter. Additionally, we modify this prior by allowing the probability of inclusion of each variable to depend on the number of its neighbors in the graph. 
{We also discuss extensions of the proposed Bayesian knockoff filter formulation to settings with non-Gaussian responses, focusing in particular on survival outcomes through accelerated failure time models.}
 
We ensure that the {latent } knockoff variables possess \textit{a priori} the required properties outlined in \cite{candes2018gold} and demonstrate that this characterization is sufficient for controlling the BFDR at an arbitrary level, in finite samples,  provided the distribution of the covariates is fully known. {When the covariate graph is estimated from data, the graphical prior yields accurate posterior estimation of the precision matrix, and the resulting procedure retains an asymptotic FDR guarantee.} In applications, to maintain control over the BFDR, we employ a greedy algorithm, similar to the one presented in \cite{newton2004bfdr}, but relying on an estimate of the upper bound for the probability of non-inclusion. 
Using simulated data, we show that our proposal substantially improves the stability of the selection compared to classical knockoff methods, as it leverages the entire posterior distribution of the knockoff variables rather than relying on a single realization. Furthermore, we show that our procedure achieves performance comparable to or better than standard Bayesian variable selection methods, while consistently controlling the FDR.

To our knowledge, only a few Bayesian knockoff filters have been proposed in the literature to date. \cite{gu2021Bayesian} first considered a generalized linear model framework and assumed the distribution of the covariates to be completely known. {In contrast, our formulation combines estimation of a Gaussian graphical model for the covariates with a parameter-expanded latent knockoff construction, so that the estimated covariate graph informs knockoff construction and inference on the regression effects. Furthermore, unlike the formulation of \cite{gu2021Bayesian}, our construction defines latent knockoff variables that constitute valid model-X knockoffs a priori.} \cite{yap2023bayesian} employ the Bayesian lasso of \cite{park2008blasso} for the estimation of the regression coefficients. However, they treat the covariates as fixed and require the augmented knockoff matrix to be generated prior to Bayesian inference.

The paper is organized as follows. Section \ref{Sec:2} provides an overview of the fundamental concepts related to knockoff filters and introduces the proposed Bayesian generalization. Section \ref{Sec:3_simulo} presents a comprehensive simulation study to evaluate the performances of the proposed method against alternative approaches. Section \ref{Sec:4_applic} uses an application to real data to demonstrate the practical utility of our approach. Some concluding remarks are presented in Section \ref{Sec:5_concludo}.

%-------------------------------------------
\section{Methods} \label{Sec:2}
% Lascio tutto, scrivo in verde quella nuova
Let $X = (X_1, \dots, X_p)$ be a set of covariates and $Y$ a response variable. We observe i.i.d. data %$(X_{i1}, \dots, X_{ip}, Y_i) \in \R^{p} \times \R$, for $i=1,\ldots,n$, and 
assembled in an $(n\times p)$ data matrix $\bm X=({x}_1, \dots, {x}_n)$ %with $i$-th row $(X_{i1}, \dots, X_{ip})$ 
and an $n$-dimensional vector $y=(y_1,\ldots,y_n)$. % with $i$-th entry $Y_i$.}
%\textcolor{red}{ Let $\bm X = (X_1, \dots, X_p)^T \in \mathbb{R}^p$ represent a vector of covariates and let $Y \in \mathbb{R}$ be the response variable. We denote the $n\times p$ matrix of observations on the covariates as $\mathcal{X} = (\bm x_1, \dots, \bm x_n)^T$ and the vector of observed responses as $\bm y = (y_1, \dots, y_n)^T$.}
In this section, we first introduce the classical notion of knockoff filters for the linear regression framework and then describe our proposed Bayesian generalization.

%-------------------------------------------------
\subsection{A review of the Knockoff filter} \label{Sec:2_KF}

Knockoff filter is a framework that enables variable selection in an FDR-controlled environment. The main idea, originally proposed by \cite{barber2015knockoff}, is to perturb the model by adding a set of inactive predictors built in such a way that they resemble the dependence structure of the original covariates. In \cite{barber2015knockoff} the knockoff predictors were built deterministically, while \cite{candes2018gold} extended the framework to the model-$X$ knockoff filter, with random covariates. The model-$X$ knockoff assumes a joint distribution $p(X)$ for the predictors and random knockoff variables, $\widetilde X=(\widetilde{X}_1, \dots, \widetilde{X}_p)$, which are defined in a way to make it hard for the model to distinguish them from $X$. Formally, the knockoffs are sampled from a distribution $p(\widetilde X \mid X)$ such that these two properties hold:
\begin{enumerate}
    \item Conditional Independence:
    \[
        \textcolor{black}{\widetilde X \ind Y \mid X}
    \]
    \item Pairwise Exchangeability:
    \[
        \textcolor{black}{\left(X, \widetilde{X}\right)_{\text{Swap}(j)} \stackrel{d}{=} \left(X, \widetilde{X}\right)} \qquad  j = 1, \dots, p,
    \]
\end{enumerate}
where \textcolor{black}{$(X, \widetilde{X})_{\text{Swap}(j)}$} results from swapping the $j$-th variable with the $j$-th knockoff. Note that while the first property is easily obtained, the second presents some difficulties \citep{sesia2018hmm, bates2021metr, berti2023new}. A general algorithm to sample knockoffs is given in \cite{candes2018gold}, but it requires the computation of the conditional probability distribution of each covariate given all the others and the previously sampled knockoffs, \textcolor{black}{$p( X_j|X_{\setminus j}, \widetilde{X}_{1:j-1})$}, which can be difficult to obtain for general distributions. On the other hand, when all covariates are pairwise independent, each knockoff can be sampled from the marginal distribution of $X_j$.

The procedure of model-X knockoffs \citep{candes2018gold} described above can be simplified if the distribution of the covariates is Gaussian. Let $X \sim N(0, \bm\Sigma)$, with $\bm\Sigma$ the covariance matrix, then a joint distribution for which pairwise exchangeability holds is
    \begin{equation}
        \label{eq:joint2}
        (X, \widetilde X) \sim N\left(0, \begin{pmatrix}
         \bm\Sigma & \bm\Sigma - \text{diag}(s)  \\
         \bm\Sigma - \text{diag}(s) & \bm\Sigma 
    \end{pmatrix} \right),
    \end{equation}
    where $s = (s_1, \dots, s_p)^T$ is a vector of values computed by minimizing the average correlation between each variable and its knockoff while ensuring that 
    \begin{equation}
    \label{eq:A}
    {\bm A}
         =2\text{diag}(s)-\text{diag}(s)\bm\Sigma^{-1}\text{diag}(s) 
    \end{equation}
    is positive definite. Thus, by the property of the multivariate distribution, the conditional distribution of the knockoffs $\widetilde X$ given the covariates $X$ is obtained as follows
    \begin{equation}
    \widetilde{X}\mid X \sim N\left(X -  X\bm\Sigma^{-1}\text{diag}(s), \bm A\right).
    \label{eq:knockoffs2}
\end{equation}

{In the classical knockoff framework as proposed by \cite{barber2015knockoff}, given the sampled knockoff variables \textcolor{black}{$\widetilde{\bm X} = (\widetilde{x}_1, \dots, \widetilde{x}_n)$}, any model which returns a variable importance metric is fit using the augmented set of predictors \textcolor{black}{$[\bm{X}, \widetilde{\bm{ X}}]$}. Variable selection is performed by defining a feature statistic $W_j$, $j=1, \ldots, p$, which combines the variable importance metric of each original predictor with the one of its own knockoff, using any antisymmetric function such that it has a symmetric distribution around zero for inactive variables. The final set of selected variables 
consists of those whose $W_j$ exceeds a pre-defined threshold, chosen according to a desired FDR level.
}

\subsection{Constructing Bayesian knockoffs}
\label{sec:bkf}
We now introduce our Bayesian generalization of the model-X knockoff filter. We are interested in the joint model for $Y, X$, i.e., $p(Y,X)$.  In particular, our interest focuses on the elements of the factorization of $p(Y,X)$ into the conditional distribution of the response $p(Y|X)$ and the marginal distribution of the covariates, $p(X)$.  For the latter we assume a multivariate Gaussian model as
\begin{equation}
\label{eq:X}
X \sim N(0, \bm\Sigma).
\end{equation}  \bcolch
Unlike the conventional model-X knockoff procedure, where a knockoff matrix is generated once and then treated as a fixed augmentation of the design matrix, in our formulation the knockoff variables are introduced as latent auxiliary variables in a parameter-expanded working model used to construct feature scores. This representation induces a paired competition between each original covariate and its matched knockoff and allows uncertainty in the knockoff layer to be propagated through posterior inference. More precisely, we take as baseline observed-data model for the response the Gaussian regression model
\begin{equation}
\label{eq:effreg}
Y \mid X, \alpha, \tau^{2} \sim N\left(X \alpha, \tau^{2} I_{n}\right),
\end{equation}
where $\alpha$ denotes the effective regression coefficient and $\tau^{2}$ the marginal error variance. To induce the competition between each original covariate and its auxiliary knockoff, we introduce a parameter expansion, such that the effective parameters $\alpha$ and $\tau^2$ are reparametrized as
\begin{equation}
\alpha=\beta+\left(I  - \bm \Sigma^{-1} \text{diag}(s)\right) \tilde{\beta},  \quad\text{and}\quad \tau^{2}=\sigma^{2}+\tilde{\beta}^{T} A \tilde{\beta},
\end{equation}
where $\beta$ and $\tilde{\beta}$ are $p$-dimensional vectors. Parameter expansion has often been used to augment a model with auxiliary parameters and latent variables while leaving the observed-data model unchanged \citep{liu1998parameter,gelman2004parameterization}. Here, we follow the same principle to obtain a reparametrization that preserves the observed-data model and underlies the construction of an antisymmetric feature statistic for variable selection. This reparametrization is motivated by the fact that it admits a convenient complete-data representation. Specifically, let \(U\) be an \(n \times p\) latent matrix with independent rows \(u_i \sim N(0,\bm A)\), where \(\bm A\) is defined as in Equation \eqref{eq:A}, and define
 \begin{equation}
         \label{eq:decomp_knock}
         \widetilde{X} =  X (I_p  - \bm \Sigma^{-1} \text{diag}(s)) + U.
\end{equation}
Then, conditional on \(X\), the latent matrix \(\widetilde{X}\) follows the Gaussian model-X knockoff distribution in Equation \eqref{eq:knockoffs2}. Equivalently, the pair \((X,\widetilde{X})\) follows the joint Gaussian distribution in Equation \eqref{eq:joint2}, and therefore preserves the usual pairwise exchangeability property. Under the reparametrization above, the baseline model in Equation \eqref{eq:effreg} admits the complete-data representation
\begin{equation}
\label{eq:completedata}
Y = X\beta + \widetilde X \widetilde\beta + \epsilon
= X\left[\beta+\left(I_p-\bm\Omega\operatorname{diag}(s)\right)\widetilde\beta\right] + U\widetilde\beta + \epsilon,
\end{equation}
where $\epsilon \sim N(0,\sigma^2 I_n)$ and $\boldsymbol{\Omega}=\boldsymbol{\Sigma}^{-1}$ is the precision matrix. Integrating out \(U\) recovers the observed-data model with effective parameters \(\alpha\) and \(\tau^2\),
\begin{equation}
\label{eq:marginalY}
Y \mid X,\beta,\widetilde{\beta},\Omega,\sigma^2
\sim
N\!\left(
X\left[\beta+\left(I_p-\Omega\operatorname{diag}(s)\right)\widetilde{\beta}\right],
\left(\sigma^2+\widetilde{\beta}^{T} A \widetilde{\beta}\right)I_n
\right),
\end{equation}
which coincides with Equation \eqref{eq:effreg} under the reparametrization above. 
Therefore, in this construction, \(\beta\) and \(\widetilde{\beta}\) should be interpreted as auxiliary components of the expanded representation of the effective coefficient \(\alpha\), rather than as two separately identifiable regression effects in the observed-data model. This expanded representation induces a paired competition between each original covariate and its matched knockoff, which we summarize through an antisymmetric feature statistic for variable selection in Section \ref{Sec:2_FDR}.

\subsubsection{Extensions to non-Gaussian responses}
{Our Bayesian knockoff filter formulation can be extended to models with non-Gaussian responses that admit a latent-variable representation for inference.
Examples include probit models \citep{albert:1993}, for binary and multinomial responses, and accelerated failure times (AFT) models \citep{kalbfleisch2002statistical, wei1992accelerated} for survival outcomes. 

 The underlying principle mirrors that of the Gaussian case.  Specifically, one begins by specifying a baseline observed-data model for \(Y\mid X\), or, when appropriate, for a latent response conditional on \(X\), and then introduces the parameter-expanded knockoff representation through the auxiliary coefficients \(\widetilde\beta\) and the latent Gaussian perturbation \(U\). This construction produces a complete-data formulation in which each original covariate is systematically paired with its knockoff analogue, while preserving the original observed-data model upon marginalization over the augmentation variables.
 
As an illustration, consider an AFT model for survival data, with censored observations. Let $T_i$ represent the time-to-event for the $i$-th subject and $c_i$ the censoring time. A Gaussian AFT model assumes $\log (T_i) = X_i \beta + \epsilon_i$, where the errors $\epsilon_i$ are i.i.d. from some error distribution. Here, we assume the errors to be i.i.d. Gaussian variables. To accommodate censored observations, we introduce a latent variable $y_i$ such that
\[
\begin{cases}
    y_i = \log(T^*_i) & \text{if } \kappa_i = 1 \\
    y_i > \log(T^*_i) & \text{if } \kappa_i = 0, \\
\end{cases}
\]
where $T^*_i=\min(T_i, c_i)$ and $\kappa_i = \mathbbm{1}{(T_i \le c_i)}$ is the censoring indicator. Conditional on \(X\), the latent response \(y_i\) then follows the same baseline Gaussian regression model as in Equation \eqref{eq:effreg}. Therefore, the parameter-expanded knockoff construction described above can be applied directly to the latent response formulation of the AFT model, leading to a corresponding complete-data representation. 

A similar argument applies to probit models through the latent-Gaussian representation of \citet{albert:1993}. More generally, the proposed framework can be adapted whenever a non-Gaussian response model can be embedded into a conditionally Gaussian latent-variable formulation. We note, however, that a general extension to arbitrary exponential-family models is more challenging. In particular, formulations that take \(p(Y\mid X,\widetilde X)\) directly as the observed-data likelihood \citep[see, e.g.,][]{gu2021Bayesian} do not align with the parameter-expanded interpretation adopted here,  because in our construction the knockoff layer is introduced only through an auxiliary representation, while the observed-data model remains \(p(Y\mid X)\). This distinction is important for preserving the intended role of the knockoffs as auxiliary variables rather than as directly observed predictors in the response model.

\ech

%We note that, even though our proposed construction extends to some type of non-Gaussian responses, a general formulation for distributions in the exponential family, like it is for the model-X Knockoff framework of \cite{candes2018gold}, is challenging to achieve within the Bayesian framework. One attempt is made in \cite{gu2021Bayesian}. However, the construction proposed by the authors does not produce valid knockoffs, since adopting $p(Y\mid X, \widetilde{X})$ as the likelihood function, as they do, implies $Y\nind \widetilde{X}, X$  and therefore the conditional independence property is not satisfied. }

\subsubsection{Sparsity inducing prior for covariates' graphical model } \label{Sec:2_pior_graph}
We regard the Gaussian distribution \eqref{eq:X} on the covariates as a graphical model and impose sparsity in the associated graph structure. Under the assumption of joint Gaussian distribution, conditional independence implies zero values in the off-diagonal elements of the precision matrix $\bm\Omega = \bm\Sigma^{-1}$.
When $\bm\Omega$ is sparse, also the matrix $\bm A$ in Equation~\eqref{eq:A} is sparse, facilitating the sampling of the knockoff variables under the pairwise exchangeability property. In addition, promoting sparsity in the precision matrix $\bm\Omega$ greatly improves computational efficiency and improves practical interpretability across various application domains. Here, we achieve this by imposing a continuous spike-and-slab prior \citep{wang2015graph} that combines a mixture prior distribution on the off-diagonal elements $\omega_{jj'}$ with an exponential distribution on the diagonal ones $\omega_{jj}$ ($j=1, \ldots, p$).  Let $\mathcal{G}$ be the undirected graph representing the conditional independences among the covariates, and let $\bm G$ be the adjacency matrix with elements $g_{jj'}\in \{0,1\}$, for $j,j'=1,\ldots,p$ and $j\ne j'$, with $1$ representing the presence of an edge between nodes $(j,j')$. Assuming the parameters in $\bm\Omega$ independent a priori, the prior distribution can be written as
\begin{equation}
\label{priorOmega}
        p\left(\bm\Omega \mid \bm G, v_0, v_1, \theta\right) = \frac{1}{K(\bm G, v_0, v_1, \theta)}\prod_{j < j'}\, N(\omega_{jj'}\mid0, v_{g_{jj'}})\,\prod_{j}\text{Exp}(\omega_{jj}\mid\theta/2), 
\end{equation}
where $K(\bm G, v_0, v_1, \theta)$ is the normalizing constant, and where $v_{g_{j,j'}}\in \{v_0,v_1\}$, with $v_0$ the variance of the Gaussian distribution for the absence of an edge and $v_1$ the variance of the edge presence.  By setting a small value for $v_0$ in the continuous spike-and-slab prior, we ensure that the entries of the precision matrix $\bm \Omega$ corresponding to non-selected edges are close to zero. 
As we will see in the next section, the graphical structure is used to inform variable selection in a way that improves performances. We complete the prior 
by assuming independent probabilities of selecting edges as 
\begin{equation}
\label{eq:prior_g}
     p(\bm G\mid v_0, v_1, \theta, \xi)  = \frac{K(\bm G, v_0, v_1, \theta)}{K(v_0, v_1, \xi)} \prod_{j<j'}\xi^{g_{jj'}}(1-\xi)^{1-g_{jj'}},
\end{equation}
where $K(v_0, v_1, \xi)$ is the normalizing constant of the distribution and $\xi$ is the prior probability of edge inclusion.

\subsubsection{Priors for variable selection}\label{Sec:2_pior_varsel}
For the regression coefficients, we adapt the spike-and-slab prior introduced in the supplementary material of \citet{candes2018gold}. Specifically, we modify their formulation by allowing each variable’s inclusion probability to depend on its number of neighbors in the estimated covariate graph, as detailed below.  {For simplicity of exposition, we formulate the prior specification and variable-selection mechanism under the augmented conditional representation in Equation \eqref{eq:completedata}; analogous arguments apply after marginalizing over the latent knockoff layer, yielding the observed-data model in Equation \eqref{eq:marginalY}.}
We first specify the joint distribution 
\begin{eqnarray}
 \label{jointprior}
		p(\beta_j, \widetilde{\beta}_j \mid h_\beta, \bm \gamma, \sigma^2) = \begin{cases}
			\delta_0(\beta_j, \widetilde{\beta}_j) & \quad \text{w.p. } \mathbb{P}[\gamma_j = 0] \\
			N(\beta_j\mid 0, h_\beta\, \sigma^2)\, \delta_0(\widetilde{\beta}_j) & \quad \text{w.p. } \mathbb{P}[\gamma_j = 1]/2 \\
			\delta_0(\beta_j)\, N(\widetilde{\beta}_j\mid 0, h_\beta\, \sigma^2) & \quad \text{w.p. } \mathbb{P}[\gamma_j = 1]/2,
		\end{cases}
	\end{eqnarray}
with $j=1, \ldots, p$.
Here, $h_\beta$ is a positive scaling hyperparameter and $\gamma = (\gamma_1, \dots, \gamma_p)^T \in \{0, 1\}^p$ is a $p$-dimensional vector of binary indicators such that $\gamma_j = 1$ { if at least one of the two coefficients, $\beta_{j}$ and $\widetilde{\beta}_{j}$, is non-zero in the expanded representation, and $\gamma_j = 0$ if both are zero}.  
By introducing two additional latent indicator vectors, $\delta = (\delta_1, \dots, \delta_p)$ and $\widetilde{\delta} = (\widetilde{\delta}_1, \dots, \widetilde{\delta}_p)$, such that $\gamma = \delta + \widetilde{\delta}$, the joint prior in Equation \eqref{jointprior} can be reformulated as a mixture of independent spike-and-slab priors on $\beta_j$ and $\tilde\beta_j$ as 
\begin{align*}
    p\left(\beta_j, \widetilde\beta_j \mid h_\beta, \delta_j, \widetilde\delta_j, \sigma^2\right) &= \delta_jN\left(\beta_j \mid 0, h_\beta\sigma^2\right)\, \delta_0(\widetilde\beta_j) +  \widetilde\delta_j\, N\left(\widetilde\beta_j \mid 0, h_\beta\sigma^2\right)\, \delta_0(\beta_j) \\
    &+ (1-\delta_j - \widetilde\delta_j)\, \delta_0(\beta_j, \widetilde\beta_j).
\end{align*}  
Allowing both coefficients to be non-zero simultaneously would weaken the intended one-to-one competition between them and make the resulting feature statistic harder to interpret. Therefore, we do not allow both indicators to assume value $1$ and choose a multinomial distribution of the type
\begin{align}
\label{eq:delta}
    \left(\delta_j, \widetilde{\delta}_j\right) \mid \gamma  & \sim\begin{cases}
        (0, 0) \quad \text{ w.p. } \mathbb{P}[\gamma_j = 0] \\
        (0, 1) \quad \text{ w.p. }  \mathbb{P}[\gamma_j = 1]/2 \\
        (1, 0) \quad \text{ w.p. }  \mathbb{P}[\gamma_j = 1]/2.
    \end{cases}
\end{align}
Furthermore, we let the probability of success depend on the graphical model on the covariates, similar to \cite{peterson2016joint}, via an Ising prior of the type
\begin{eqnarray}
\label{isingprior}
    p(\gamma \mid \bm G, a, b) \propto \exp\left(a\mathbbm{1}^T \gamma + b\gamma^T \bm G \gamma\right),
  \end{eqnarray}
where $\bm G$ is the adjacency matrix of the undirected graph on the covariates. The hyperparameter $a$ controls the overall sparsity of the vector $\gamma$, while $b$ is linked to the graph influence on the selection. Specifically, increasing the value of $b$ increases the probability to have $\gamma_j = 1$ for the variables with many edges in the graph. Lastly, we complete our prior model with a conjugate prior on the variance $\sigma^2$,
\begin{equation}
\label{eq:sigma_prior}    
\sigma^2 \sim \text{IG}(a_\sigma, b_\sigma).
\end{equation}
We note that \cite{candes2018gold} adopt prior in Equation~\eqref{jointprior} in a fixed design Bayesian regression with augmented covariates matrix $[\bm X, \widetilde{\bm X}]$ where $\widetilde{\bm X}$ is sampled once from Equation~\eqref{eq:knockoffs2}. On the contrary, we consider the knockoffs as latent variables and update them in the posterior sampling along with the other parameters.

%----------------------------------------------
\subsubsection{Markov Chain Monte Carlo for posterior sampling} 
\label{Sec:2_MCMC}

Let $\bm D = \left\{x_i, y_i\right\}_{i=1}^n$ represent the sample data and let $\Theta$ indicate the set of unknown model parameters and latent variables, $\Theta = \{\beta, \widetilde{\beta}, \sigma^2, \delta, \widetilde{\delta}, \gamma, \bm\Omega, \bm G\}$. 
The joint posterior distribution is proportional to the product between the likelihood and the prior distributions
\[
\begin{split}
    p(\Theta \mid  \bm D) 
    &\propto{} p(Y\mid X,\Theta) \; p(X \mid \Theta)\; p(\Theta)\\ 
    &\propto{}  p(Y\mid X, \beta, \widetilde{\beta}, \bm\Omega, \sigma^2) \, p(X \mid \bm\Omega) \, p(\beta, \widetilde{\beta}\mid\delta, \widetilde{\delta}) \, p(\delta, \widetilde{\delta}\mid \gamma)\\
    & \times p(\bm\Omega\mid \bm G)\, p(\gamma\mid \bm G)\, p(\bm G)p(\sigma^2).\\
\end{split}
\]
\noindent This distribution is not analytically tractable, and we therefore resort to a Markov Chain Monte Carlo (MCMC) algorithm \citep{brooks2011handbook}. In particular, we implement an efficient Metropolis within Gibbs scheme that employs a data augmentation (DA) scheme based on the latent variable $U$ introduced in equation \eqref{eq:decomp_knock} and the  { complete-data representation} of our model detailed in Section \ref{sec:bkf}. { Specifically,  since it is not possible to obtain tractable full conditional distributions under the marginal model \(p(Y\mid X,\beta,\tilde{\beta},\Omega,\sigma^2)\), we work with the augmented representation \(p(Y\mid X,U,\beta,\tilde{\beta},\Omega,\sigma^2)\) and then update \(U\) from its full conditional distribution.} We also incorporate the Add-Delete Metropolis-Hastings algorithm of \cite{savitsky2011variable}, that cleverly avoids having to deal with the changing dimensions of the parameter space via a joint update of the inclusion indicators of the spike-and-slab priors and the corresponding regression coefficients.  We provide full details of the MCMC algorithm in the Supplementary Material. Briefly, our algorithm is composed by the following steps:

\begin{enumerate}
    \item We jointly update $(\delta, \widetilde{\delta}, \gamma, \beta, \widetilde{\beta})$ with a stochastic search approach. At each iteration $t = 1, \dots, T$, we randomly select one $j \in \{1, \dots, p\}$ and propose to change the value of $\gamma_j$. If the proposed $\gamma_j^* = 0$ then the only possible choice for the proposed $\beta_j^*$,  $\widetilde\beta_j^*$, $\delta_j^*$, $\widetilde\delta_j^*$ is to set them all to $0$. However, if we select a non-included variable, $\gamma_j = 0$, we randomly propose to include the original variable or its knockoff. More specifically, we propose the selected coefficient using a Gaussian distribution $N(0, 0.5)$ as the proposal distribution. We then accept the proposed parameters with probability $r=\min\left\{1, \rho\right\}$ with
            \[
                \rho = \frac{p(Y\mid X, U, \beta^*, \widetilde\beta^*, \bm\Omega, \sigma^2)p(\beta^*, \widetilde\beta^* \mid \gamma^*, \sigma^2, h_\beta)p(\delta^*, \widetilde\delta^* \mid \gamma^*)p(\gamma^* \mid \bm G, a, b)}{p(Y\mid X, U, \beta, \widetilde\beta, \bm\Omega, \sigma^2)p(\beta, \widetilde\beta \mid \gamma, \sigma^2, h_\beta)p(\delta, \widetilde\delta \mid \gamma) p(\gamma \mid \bm G, a, b)}
            \]
    
%    \textcolor{teal}{where $\widetilde{\bm X}^{(t)} = \bm X (I  - \bm \Omega^{(t)} \text{diag}(s)) + U^{(t)}$}.

To improve the mixing of the algorithm, we perform a within-model update step, proposing a new value for each selected coefficient using as proposal a distribution centered on the current value of $(\beta, \widetilde{\beta})$.

\item We sample $\sigma^2$ from its full conditional distribution 
% \[
%     \sigma^2 \mid y, \bm X, U, \beta, \widetilde{\beta}, \bm\Omega \sim IG[a_\sigma+\frac{n}{2}, b_\sigma+\frac{\sum_i (y_i - x_i(\beta + (I - \bm\Omega\text{diag}(s))\widetilde\beta) - U_i\widetilde\beta)^2}{2}]
% \]
\[
\begin{aligned}
\sigma^2 \mid y,\bm X,U,\beta,\widetilde\beta,\bm \Omega
\sim\;
IG\Biggl(
& a_\sigma+\frac n2, \\
& b_\sigma+\frac12\sum_{i=1}^n
\Bigl(
y_i
- x_i^T\bigl[\beta+\bigl(I-\bm \Omega\operatorname{diag}(s)\bigr)\widetilde\beta\bigr]
- u_i^T\widetilde\beta
\Bigr)^2
\Biggr).
\end{aligned}
\]
% \[
% \sigma^2 \mid y,\bm X,U,\beta,\widetilde\beta,\bm \Omega
% \sim
% IG\!\left(
% a_\sigma+\frac n2,\;
% b_\sigma+\frac12\sum_{i=1}^n
% \left(
% y_i-x_i^T\left[\beta+\left(I-\bm \Omega\operatorname{diag}(s)\right)\widetilde\beta\right]-u_i^T\beta^e
% \right)^2
% \right).
% \]

\item  We update $(G, \bm \Omega)$ from $p(G, \bm \Omega \mid X)$ via a block Gibbs sampler similar to the one used by \cite{wang2015graph}. 
\item 
Lastly, we update the rows \(u_i\) of the latent auxiliary variable \(U\) from their full conditional distributions
\[
u_i \mid y,X,\beta,\widetilde\beta,\Omega,\sigma^2 \sim N(\mu_i,\Sigma_U),
\]
where
\[
\Sigma_U = \left(A^{-1}+\frac{1}{\sigma^2}\widetilde\beta(\widetilde\beta)^T\right)^{-1},
\]
and
\[
\mu_i = \Sigma_U\frac{1}{\sigma^2}\widetilde\beta
\left(
y_i - x_i^T\left[\beta+\left(I-\Omega\operatorname{diag}(s)\right)\widetilde\beta\right]
\right),
\qquad i=1,\dots,n.
\]
% Lastly, we update the elements of the latent auxiliary variable $U$ from their full conditional distributions
% \[
%     u_i  \mid y, \bm X, U, \beta, \widetilde\beta, \bm\Omega, \sigma^2 \sim N(\mu_i, {\bm \Sigma}_U)
% \]
% \[
%         {\bm \Sigma}_U = \left({\bm A} + \frac{1}{\sigma^2}\widetilde{\beta}\,\widetilde{\beta}^T\right)^{-1}
% \]
% \[
%         \mu_i = {\bm \Sigma}_U \left(\frac{1}{\sigma^2}\left( y_i\widetilde{\beta} - \widetilde{\beta}\, \beta^Tx_i -\widetilde{\beta}\widetilde{\beta}^T\Gamma_i\right)\right)
% \]
% \[
%     \Gamma = X(I-{\bm \Omega}\text{diag}(s)),
% \]
% for $i = 1, \dots, n$.

\end{enumerate}

\subsection{Controlling the false discovery rate} \label{Sec:2_FDR}
In this section, we demonstrate that our model effectively performs variable selection while controlling the Bayesian FDR. In Bayesian frameworks, the conventional definition of the false discovery rate is not directly applicable, as both the set of active and inactive variables are treated as random variables. To address this, it is customary to rely on a notion of Bayesian FDR, as discussed for example by \cite{whittemore2007bfdr}, where it is defined as the expected value of the false discovery proportion conditioned on the observed data, $\bm{D}$, as
\begin{equation}
    \label{eq:bfdr}
        BFDR(\mathcal{S}) = \mathbb{E}\left[\frac{\vert \{ j \in \mathcal{S} : X_j\ind Y {\mid X_{-j}}\} \vert }{\vert  \mathcal{S} \vert \lor 1} \; \bigg| \; \bm  D \right],
\end{equation}
where $\mathcal{S}\subseteq\{1, \dots, p\}$ is the subset of selected indexes. %Here, the expectation is taken with respect to the posterior distribution of the selection rule, given the observed data. 
Thus, the BFDR represents the expected proportion of incorrect rejections among all possible rejections, where the expectation is conditioned on the posterior distribution of active and inactive variables given the observed data. Proposition \ref{th:1} states that, given an antisymmetric function $W_j$, our proposed method controls the BFDR at level $q$.
\begin{proposition}
\label{th:1}
  \bcolch   Let the latent knockoff matrix \(\tilde X\) be induced from the augmentation variable \(U\) through Equation \eqref{eq:decomp_knock}, and assume that the resulting pair \((X,\tilde X)\) is pairwise exchangeable, i.e.\ech
    %Let the prior distribution for the knockoff latent variable  $U$ be such that pairwise exchangeability holds, i.e.
    \[
        p(X, \widetilde{X})_{\text{Swap}(j)} = p(X, \widetilde{X}) \qquad \forall j \in \{1, \dots, p\}.
    \]
\bcolch 
Assume further that the response follows the complete-data representation in Equation \eqref{eq:completedata}, with prior distribution on \((\beta,\tilde\beta)\) invariant to swaps,\ech
% Let the distribution of the response variable $Y$ follow the model in Equation \eqref{eq:pY} with prior distribution on the regression coefficient invariant to swaps, i.e.
    \[
        p(\beta, \widetilde{\beta})_{\text{Swap(j)}} = p(\beta, \widetilde{\beta}) \quad j \in \{1, \dots, p\}.
    \]
    Then, the set  $\widehat{\mathcal{S}} = \argmax_{\mathcal{S} \subseteq \{1, \dots, p\}} \vert \mathcal{S} \vert$  subject to 
    %\textcolor{teal}{AG: farei un remind a cosa è $W_j$. Inoltre, l'ipotesi sulla sign-flip credo vada messa qui, anche semplicemente a parole, e non nella proof. Forse.} \textcolor{red}{LF: Il remind l'ho messo. La sign-flip credo vada nella proof perché è consegienza delle ipotesi (l'hanno dimostrato gu e yin.}

    \[
      \frac{1}{\vert\mathcal{S}\vert \lor 1}\sum_{j \in \mathcal{S}}\left\{ 1 - \mathbb{P}\left [W_j > 0 \mid \bm D \right ] + \mathbb{P}\left [W_j < 0 \mid \bm D \right ] \right\}\le q,
    \]
    where $W(\beta, \widetilde{\beta})$ is { any} antisymmetric function, controls the $BFDR$ at level $q$.
\end{proposition}
\begin{proof}
        Let $r = (r_1, \dots, r_p) \in \{0, 1\}^p$ be a random vector such that $r_j = 1 \iff X_j\nind Y \mid X_{-j}$ with $X_{-j} = (X_1, \dots, X_{j-1}, X_{j+1}, \dots, X_p)$. Under a compound loss function \citep{muller2006fdr}, the Bayesian FDR can be written as
    \[
        BFDR(\mathcal{S}) = \frac{\sum_{j \in \mathcal{S}}\mathbb{P}[r_j = 0 \mid \bm D]}{\vert \mathcal{S}\vert \lor 1},
    \]
    where $\mathbb{P}[r_j = 0 \mid \bm D]$ represents the marginal posterior probability of a covariate to be inactive. Assume for now that the random vector $W$ satisfies the sign-flip property, i.e.
    \[
        \mathbb{P}[W_j > 0 \mid \bm D, r_j = 0] = \mathbb{P}[W_j < 0 \mid \bm D, r_j = 0].
    \]
    %{\color{black} Moreover, by construction, we have that $\mathbb{P}[W_j > 0 \mid \bm D, r_j = 1] > \mathbb{P}[W_j < 0 \mid \bm D, r_j = 1]$. Therefore, marginally with respect to $r_j$, we obtain
    %\[
    %    \mathbb{P}[W_j > 0 \mid \bm D] > \mathbb{P}[W_j < 0 \mid \bm D].
    %\]
    %}
    By the law of total probability, we can write

    \begin{align*}
         \mathbb{P}[W_j > 0\mid \bm D] & = \mathbb{P}[W_j > 0\mid \bm D, r_j = 0]\mathbb{P}[r_j = 0\mid \bm D]\\
         & +  \mathbb{P}[W_j > 0\mid \bm D, r_j = 1]\mathbb{P}[r_j = 1 \mid \bm D].
\end{align*}

\noindent
Applying the same argument to $\mathbb{P}[W_j < 0 \mid \bm D]$ and using the sign-flip property of $W$, we obtain
    \begin{align*}
        \mathbb{P}[W_j > 0\mid \bm D] - \mathbb{P}[W_j < 0\mid \bm D] & = \mathbb{P}[r_j = 1\mid \bm D]\times\left(\mathbb{P}[W_j > 0 \mid \bm D, r_j=1]\right.\\
        & \left.- \mathbb{P}[W_j < 0 \mid \bm D, r_j = 1]\right).
    \end{align*}
    Then, we can conclude,
	\begin{equation} \label{eq:inequality_proof}
	   \mathbb{P}[W_j > 0\mid \bm D] - \mathbb{P}[W_j < 0\mid \bm D] \; \le \;  \mathbb{P}[r_j = 1\mid \bm D]. 
	\end{equation} 
 Since $\mathbb{P}\left[r_j = 0\mid \bm D\right] \, = \, 1-\mathbb{P}[r_j=1\mid \bm D]$, the inequality in Equation~\eqref{eq:inequality_proof} provides an upper bound for $\mathbb{P}\left[r_j = 0\mid \bm D\right]$, i.e.,
	\begin{align}
		\mathbb{P}\left[r_j = 0\mid \bm D\right] \le 1 - \mathbb{P}[W_j > 0\mid \bm D] + \mathbb{P}[W_j < 0\mid \bm D].
  \label{ubound}
\end{align}
To complete the proof, we need to show that an invariant prior on $(\beta, \widetilde{\beta})$ leads to a $W$ that satisfies the sign-flip property. This result can be obtained using a similar argument as in Theorem~1 of \cite{gu2021Bayesian}. We report the proof in the Supplementary Material for completeness.
\end{proof}

In practice, we can compute the upper bound for $\mathbb{P}\left[r_j = 0\mid \bm D\right]$ in Equation \eqref{ubound} via Monte Carlo sampling given the MCMC output, as
\begin{equation}
    \label{eq:estimate_ubound}
    \widehat{\mathbb{P}}\left[r_j = 0\mid \bm D\right] \le 1 - \frac{1}{T}\left(\sum_{t=1}^T \mathbbm{1}_{\{W_j^{(t)} > 0\}} - \mathbbm{1}_{\{W_j^{(t)} < 0\}} \right),
\end{equation}
where $T$ is the number of posterior samples. A simple algorithm to construct the set $\widehat{\mathcal{S}}$ required by Proposition \ref{th:1} is to compute the upper bound in Equation \eqref{eq:estimate_ubound} for each covariate and sort them such that the bounds are in ascending order. Then, add variables to the set until the average of the upper bounds is below $q$. {Proposition \ref{th:1} only requires $W_j$ to be antisymmetric. In practice, a choice of $W_j$ is needed to use the model. Many proposals are available in the literature \citep{barber2015knockoff, spector2022powerful}. We adopt the simplest function
\begin{equation*}
    W_j= \Big\vert \beta_j\Big\vert - \left\vert \widetilde{\beta_j}\right\vert
\end{equation*}
in all applications of this paper, since we found that the final selection is not sensitive to the choice of $W_j$.}

We remark that our model performs variable selection using an upper bound on the  posterior probability of non-inclusion. This strategy allows us to overcome issues with strong correlation structure among the covariates, which typically affect the state-of-art variable selection methods that rely on the estimated marginal posterior probabilities of inclusion, as empirically shown in our simulation study. Our theoretical results, indeed, show that the bound used for our variable selection procedure is not affected by the dependency structure of the covariates.

Although Proposition \ref{th:1} shows that our  method controls the BFDR at a threshold $q$, it assumes that the latent variables $\widetilde{X}$ are  valid knockoffs a priori. As discussed by \cite{candes2018gold}, the construction of valid model-X knockoffs requires perfect knowledge of the covariates' distribution. However, our proposed model assumes that the covariates $X$ follow a Gaussian distribution, with an  unknown precision matrix that must be estimated from the data. \cite{barber2020rob} prove that the knockoff procedure maintains FDR control at a desired threshold $q$ if the precision matrix of the Gaussian model is correctly estimated. Here, we show that, under mild regularity conditions, the priors chosen to estimate the Gaussian graphical model for the covariates yield accurate posterior estimates of the precision matrix. %Thus, drawing on similar arguments as in \cite{barber2020rob}, we can conclude that our procedure controls both the frequentist FDR and, consequently, the expected BFDR \citep{guindani2009bayesian} in large samples, even if the precision matrix is unknown. 
More precisely, we are able to prove that the posterior convergence rate of our estimator is the same as the one of the Bayesian graphical lasso. To show such result, we first need to prove that our prior distribution assigns enough probability mass to values ``close'' to the true one.
\begin{proposition}
\label{prop:prior_rate}
    Let  the true precision matrix ${\bm \Omega}_0 \in \mathcal{U}(\epsilon_0, l) = \{{\bm \Omega} : \vert\{(i, j) : 1 \le i < j \le p, \omega_{ij} \ne 0\}\vert \le l, 0 < \epsilon_0 \le \text{eig}({\bm \Omega})_1 < \text{eig}({\bm \Omega})_p \le \epsilon_0^{-1} < \infty \}$ for some $0 < \epsilon_0 < \infty$ and $0 \le l \le p(1-p)/2$. Also assume that the prior distributions for the precision matrix $\Omega$ and the adjacency matrix $\bm G$ are:
$$
p\left(\bm\Omega \mid \bm G, v_0, v_1, \theta\right) = \frac{1}{K(\bm G, v_0, v_1, \theta)}\prod_{j < j'}N(\omega_{jj'}\mid0, v_{g_{jj'}})\prod_{j}\text{Exp}(\omega_{jj}\mid\theta/2)
$$
with $v_{g_{ij}} \le \frac{\Vert{\bm \Omega}_{0}\Vert_\infty}{2\log 2}$ and
$$
p(\bm G\mid v_0, v_1, \theta, \xi)  = \frac{K(\bm G, v_0, v_1, \theta)}{K(v_0, v_1, \xi)} \prod_{j<j'}\xi^{g_{jj'}}(1-\xi)^{1-g_{jj'}},
$$
with $\xi < 0.5$. Then, 
$$
    \Pi(B(p_{\Omega_0}, \epsilon_n)) \gtrsim (c\epsilon_n/p)^{p+l},
$$
with $\Pi(\cdot)$ the probability under the prior distribution, $B(p_{\Omega_0}, \epsilon_n) = \{p: K(p_{\Omega_0}, p_{\Omega}) \le \epsilon_n^2, V(p_{\Omega_0}, p_{\Omega}) \le \epsilon_n^2\}$ with  $K(f, g) = \int f\log(f/g)$ and $V(f, g) = \int f\log^2(f/g)$ and $p_{\Omega_0}$ and $p_\Omega$ the density functions of $N(0, {\bm \Omega}_0)$ and $N(0, {\bm \Omega})$ respectively. 
\end{proposition}
\begin{proof}
    See the Supplementary Material.
\end{proof}

Using the prior concentration result stated in Proposition \ref{prop:prior_rate}, we can investigate the posterior convergence rate of the graphical parameters. Proposition \ref{prop:post} shows how the posterior convergence rate of our model is equal to the convergence rate of the graphical lasso \citep{friedman2008glasso} and the Bayesian graphical lasso \citep{park2008blasso}. Note that, in Proposition \ref{prop:post}, the prior distribution for the precision matrix in Equation \eqref{eq:prior_bound} is slightly different from the one used in our model in Equation \eqref{eq:prior_g} since it includes the probability $\mathbb{P}[\bar{R} \ge \vert \bm G \vert]$ to penalize denser graphs. However, in the context of MCMC algorithm to sample from posterior distribution, such probability has little to no effect. The ratio of these probabilities is, indeed, close or superior to 1 
%\textcolor{teal}{AG: ci va bene anche superior?}{\color{red}LF: sì, il rapporto di accettazione si definisce come il minimo tra 1 e quel ratio.} 
for most proposal distributions.
\begin{proposition}
\label{prop:post}
Let $X^{(n)} = (X_1, \dots, X_n)$ be a random sample from the p-dimensional Gaussian distribution with mean $0$ and precision matrix ${\bm \Omega}_0 \in \mathcal{U}(\epsilon_0, l) = \{{\bm \Omega} : \vert\{(i, j) : 1 \le i < j \le p, \omega_{ij} \ne 0\}\vert \le l, 0 < \epsilon_0 \le \text{eig}({\bm \Omega})_1 < \text{eig}({\bm \Omega})_p \le \epsilon_0^{-1} < \infty \}$ for some $0 < \epsilon_0 < \infty$ and $0 \le l \le p(1-p)/2$. Also assume that the prior distributions are:
$$
p\left(\bm\Omega \mid \bm G, v_0, v_1, \theta\right) = \frac{1}{K(\bm G, v_0, v_1, \theta)}\prod_{j < j'}N(\omega_{jj'}\mid0, v_{g_{jj'}})\prod_{j}\text{Exp}(\omega_{jj}\mid\theta/2)
$$
with $v_{g_{ij}} \le \frac{\Vert{\bm \Omega}_{0}\Vert_\infty}{2\log 2}$ and
\begin{equation}
\label{eq:prior_bound}
p(\bm G\mid v_0, v_1, \theta, \xi)  = \frac{K(\bm G, v_0, v_1, \theta)}{K(v_0, v_1, \xi)} \prod_{j<j'}\xi^{g_{jj'}}(1-\xi)^{1-g_{jj'}} \mathbb{P}(\bar{R} \ge \vert \bm G \vert),    
\end{equation}
with $\xi < 0.5$ and $\mathbb{P}(\bar R > a_1m) \le \exp\{-a_2m\log m\}$ for some constants $a_1$ and $a_2$ and $m = 1, 2, \dots$. Then, 
$$
\mathbb{E}_0 [\mathbb{P}\{\Vert {\bm \Omega} - {\bm \Omega}_0\Vert_2 > M\epsilon_n \mid X^{(n)}\}] \to 0
$$
for $\epsilon_n = n^{-1/2}(p+l)^{1/2}(\log p)^{1/2}$ and sufficiently large constant $M$.
\end{proposition}
\begin{proof}
    See the Supplementary Material.
\end{proof}

\noindent Proposition \ref{prop:post} shows that, with sufficiently small prior variances, the posterior mean of the precision matrix ${\bm \Omega}$ converges to the true one with the same rate as the convergence rate of graphical lasso \citep{rothman_sparse_2008}.  Therefore, employing similar arguments as in \cite{barber2020rob}, we can show that our model controls the
frequentist FDR up to an inflation factor that depends on how well the conditional distribution of $X$
(and hence the knockoff mechanism) is approximated. More specifically, we establish that, for a fixed  nominal target level $q\in (0, 1)$, the frequentist
FDR is controlled at level $q e^{\eta_n}$, where $\eta_n$ depends on the posterior contraction rate of
${\bm \Omega}$ around the true precision matrix ${\bm \Omega}_0$ and the sparsity of ${\bm \Omega}_0$.
In high-dimensional settings, only an inflated FDR control can be achieved under the conditions of
Proposition \ref{prop:post}, similarly as in \cite{barber2020rob}. Tighter control may be possible under additional
structural assumptions (e.g., bounded maximum degree together with $\ell_\infty$ contraction of the posterior). 
More specifically, we can prove the following result:

\begin{proposition}
\label{thm:fdr-bcs-clean}
Let $X_{1}, \ldots, X_{n} \stackrel{i i d}{\sim} N_{p}\left(0, \Omega_{0}^{-1}\right)$ with $\Omega_{0} \succ 0$,
and assume that $\lambda_{\min}(\Omega_0)\ge m_0>0$.
Assume also that $\log p=o(n)$ and that Proposition 3 holds with contraction rate
$\varepsilon_n$, with $\varepsilon_n\to 0$.
Let $\widehat S$ be as in Proposition 1, and let
\[
\E^*[\cdot]
:= \E_{X^{(n)} \sim \Pbb_0}\,
   \E_{\Omega \sim \Pi(\cdot \mid X^{(n)})}\,
   \E_{\tilde{X}\mid \Omega, X^{(n)}}[\cdot],
\]
denote the marginal expectation with respect to the joint law of the $X$'s, the posterior of $\Omega$,
and the knockoff generation. Then there exists a sufficiently large constant \(M>0\) such that
\[
\E^*\!\left[\mathrm{FDP}(\widehat S)\right]
\;\le\; q\,e^{\eta_n} + o(1),
\]
where $q\in (0,1)$ and
\(
\eta_{n}
:=\Big(2M^{2} n\varepsilon_{n}^{2}
+4M \varepsilon_{n} \sqrt{n \log p}\Big)(1+o(1)).
\)
\end{proposition}
}

%---------------------------------------------- 
\section{Simulation study}\label{Sec:3_simulo}
We evaluate the performance of the proposed model through a comprehensive simulation study and compare results with those of the classical model-X knockoff filter and state-of-the-art variable selection methods, including lasso regression \citep{tibshirani1996lasso} and models that use discrete spike-and-slab priors \citep{mitchell1988ssl}.

\subsection{Data generation}
\label{sec:sim_gen}
\begin{figure}
     \centering
     \includegraphics[width=\textwidth]{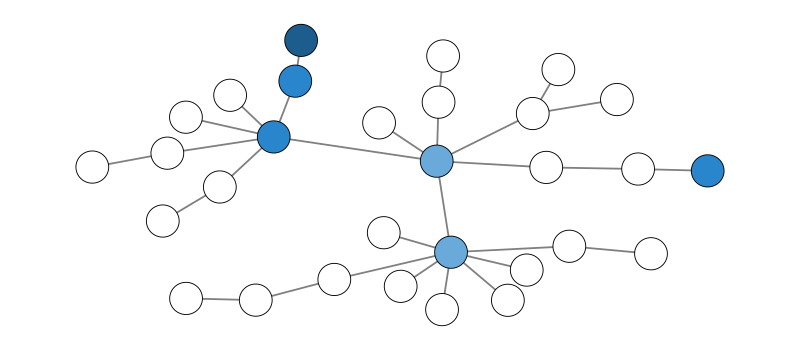}
     \caption{{\bf Simulation study:} Scale-free graph structure under Gaussian graphical model in Scenario $2$. White nodes correspond to the noise variables, and black nodes to the active ones. A stronger black color corresponds to a larger effect.}
     \label{fig:graph}
 \end{figure}

We generated data from a few distinct scenarios. In Scenario $1$, we generated $p=30$ covariates independently sampled from a standard Gaussian distribution. Out of the $30$ covariates, we assumed that $6$ of them have a linear effect on the response variable $y$. The regression coefficients associated with these active covariates were randomly selected from the set $\{\pm 0.5, \pm 1, \pm 1.5\}$ to introduce variation in the strength of the effects.  We then simulated the response variable from the linear equation  $y_i =  x_i\beta + \epsilon_i$, with $\epsilon_i \sim N(0, \sigma_\epsilon)$ and $\sigma_\epsilon=1$, for $i=1,\ldots,n$. This relatively simple scenario enables us to assess the capability of our model to recover the set of active variables when the graphical structure learning is, effectively, a source of noise in the model. 
In Scenario $2$, we introduce a more realistic setting by incorporating a correlation structure among the predictors. More specifically, the $30$ covariates were generated from a Gaussian graphical model, where the underlying graph structure is characterized by a sparse precision matrix. The underlying graph structure is shown in Figure~\ref{fig:graph}. Similarly to Scenario $1$, we selected $6$ out of the $30$ variables to be active in explaining the response $y$ and randomly sampled their coefficient values from the set $\{\pm 0.5, \pm 1\}$. 
%For these two scenarios we set $n=200$. Scenario $3$, has the same graph and active variables as in Scenario $2$, but with a smaller sample size ($n=25$) and larger active variable coefficients sampled from $\{\pm 2\}$.
%active variables, whose regression coefficient was set to be equal to $2$, to create a stronger effect. %$\beta_j = 2$. 
In Scenario $3$, data are simulated following the setting of \cite{li2008network} and \cite{peterson2016joint}. Specifically, we generated the covariates from a graph with 40 hubs, each with 5 connected nodes. This resulted in $240$ covariates with a sparse graphical model of $200$ edges in total. We set as active variables those in the first $4$ hubs, plus their connected edges, resulting in a total of $24$ non-zero coefficients. In particular the regression coefficients of the 4 hubs are $5$, $-5$, $3$ and $-3$, respectively, while the coefficients of the associated nodes are the same but divided by $\sqrt{10}$. The covariates were then generated from a Gaussian distribution with zero mean and covariance matrix $\bm\Sigma$ with diagonal elements equal to $1$ and covariances between connected variables equal to $0.7$. As in the previous scenarios, the response variable was simulated from a linear Gaussian model with the error variance equal to $\sum_j\beta_j^2 / 4$. In each scenario we set the sample size at $n = 200$ and considered $100$ replicated data sets. 
 
\subsection{Parameter settings}
When fitting our model, we need to specify the hyperparameters of the shrinkage inducing prior  \eqref{priorOmega} on the precision matrix of the covariates' graph, the spike-and-slab priors \eqref{jointprior} on the regression coefficients and the Ising prior \eqref{isingprior} on the selection indicators. In all analyses, for prior \eqref{priorOmega} on the precision matrix, following the recommendations in \cite{wang2015graph}, we used $v_0 = 0.01^2$ and $v_1 = h v_0$ with $h = 100$. We set the prior probability of edge selection $\xi$ to $0.01$ and the exponential parameter $\theta$ for the elements on the diagonal to $2$. To specify vaguely informative priors, we set $h_\beta=1$ in the joint prior \eqref{jointprior} on the regression coefficients, and  $\alpha_{\sigma}=\beta_{\sigma}=2$ for the prior \eqref{eq:sigma_prior} on the error variance. Furthermore,
we centered covariates and response variables, ensuring that the intercept term in the model is zero. \cite{smith1973int} showed that, in the Bayesian framework, setting the intercept term to zero is equivalent to placing a Gaussian prior with infinite variance on the intercept parameter. Finally, for the Ising prior \eqref{isingprior} on $\gamma$,  we set $a = b = 0.5$. This specification corresponds to a prior inclusion probability of $0.622$ on the original variable or its knockoff, when the graph is empty. When experimenting with this prior, we found non-sparse specifications to work well, as the Bayesian knockoff framework allows for further control of the FDR by excluding inactive covariates. We provide sensitivity analyses to these prior choices in Section \ref{sec:sens}.

When initializing the MCMC algorithm, we set the selection indicators $\gamma$ to $0$ and assumed independent covariates with $\bm \Omega  =\bm I$. The results reported below in Section \ref{sub:results} are obtained by running MCMC chains with $8,000$ iterations for burn-in, followed by $8,000$ iterations for inference. For each chain, we assessed convergence by using the Geweke's diagnostic test \citep{geweke1992conv} to check for signs of non-convergence. For example, the average $z$-scores from the test for $\{(\beta_j, \widetilde\beta_j)\}_{j=1}^p$ were $0.03$, $0.01$ and $ -0.05$ for Scenario $1$, Scenario $2$ and Scenario $3$, respectively. This suggests that the algorithm has run for enough iterations.

\subsection{Results and comparisons} \label{sub:results}
\begin{figure}
    \centering
    \setkeys{Gin}{width=\textwidth, keepaspectratio}
    \begin{subfigure}{\textwidth}
        \includegraphics[width=\textwidth]{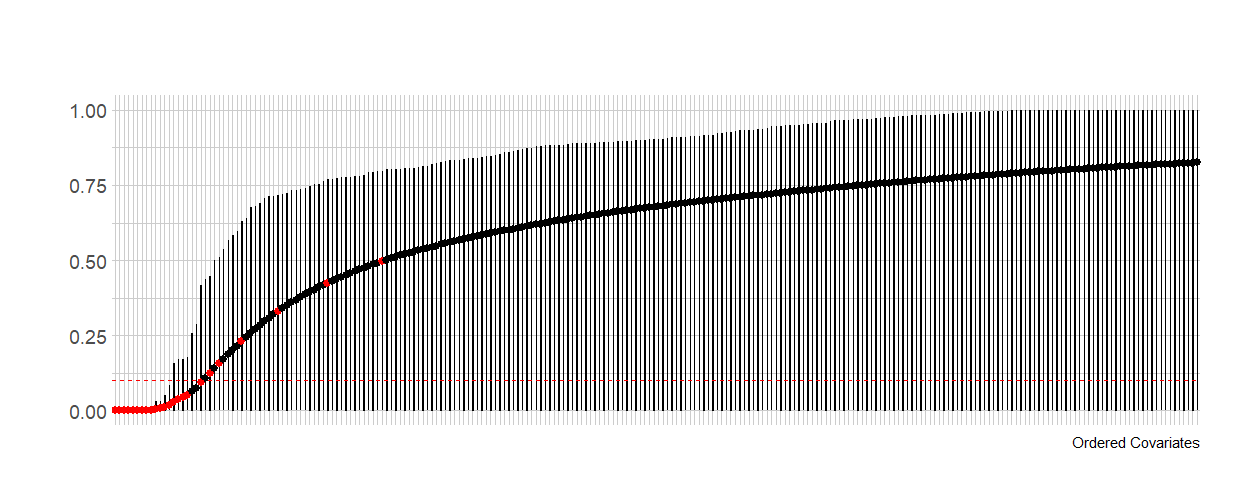}
        \label{fig:comparison_a}
    \end{subfigure}
    \begin{subfigure}{\textwidth}
        \includegraphics[width=\textwidth]{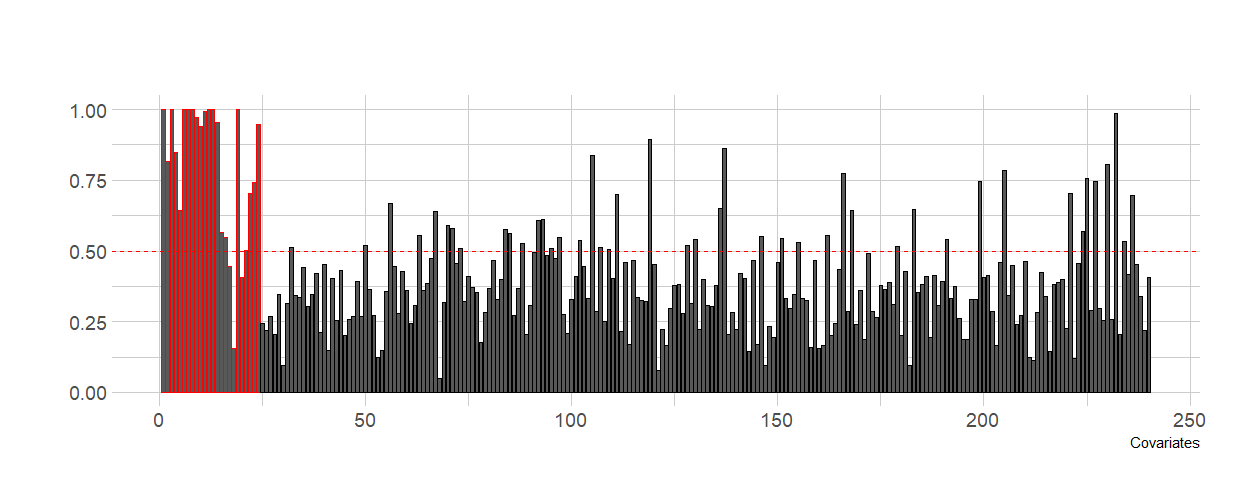} 
        \label{fig:comparison_b}
    \end{subfigure}
    \caption{{\bf Simulation study:} Variable selection performed by our Bayesian knockoff filter method (BayeKnock) and by spike-and-regression (Spike-Slab) on a simulated dataset under Scenario 3. {\it Top plot (BayesKnock)}: Bars represent estimates of the upper bound for ${\mathbb{P}}[r_j=0\mid D]$ as in Equation \eqref{ubound} while points represent the estimated BFDR. Red points correspond to true active variables. {\it Bottom plot (Spike-Slab)}: Estimated marginal posterior probabilities of inclusion (PPIs). Red bars correspond to true active variables.}    \label{fig:comparison}
\end{figure}

Let us first examine a single data set to gain insight into how the proposed model works. Specifically, we focus on a randomly selected data set generated under Scenario 3. The top plot of Figure \ref{fig:comparison} shows the estimated BFDR and the upper bound for {$\mathbb{P}[r_j=0\mid \bm D]$}, for $j=1,\ldots p$ and covariates sorted by the BFDR values. Our method correctly identifies 18 of the 24 true predictors with 6 false negatives, for a $q=0.1$ threshold on the BFDR. The bottom plot of Figure \ref{fig:comparison} reports the PPIs of all covariates, calculated by fitting a traditional Bayesian linear regression model with discrete spike-and-slab priors and a Beta hyperprior on the inclusion probability of the latent indicators \citep{george1997bvs, brownetal98, vannucci2021discrete}. This method selects 21 of the 24 true predictors with 44 false positives, for a $0.5$ threshold on the PPIs. The large number of false positives is due to the fact that the presence of collinearity in the data greatly affects the PPIs values, resulting in inflated values for inactive covariates that are correlated to the active ones. This simple example highlights the inability of clearly dividing active and inactive variables based on the posterior probability values. The results are confirmed in the comparison study below, that shows performance values averaged over replicated datasets. This empirical comparison of our Bayesian Knockoff with a traditional spike and slab regression confirms the theoretical results in Proposition \ref{th:1}, where the upper bound is unaffected by the covariates' dependency structure.

%Spike e slab: FDR = 0.6769231, TPR = 0.875
%Noi: FDR =  0.1, TPR = 0.75
%Noi: TP = 18 su 24, TN = 214 su 216, FP = 2 su 20, FN = 6
%Spike and Slab: TP = 21 su 24, TN = 172 su 216, FP =  44 su 65, FN =  3

Next, we report results of our method averaged over 100 replicated datasets and perform a comparison of the performances with other procedures for variable selection, with and without FDR control. 
Among these procedures, we include the classical model-X knockoff filter proposed by \cite{barber2019knockoff}. In particular, we consider two versions of the model-X knockoff filter, one with the true covariance matrix assumed known, and one with an estimate of $\bm\Sigma$ obtained via the graphical lasso estimator of \cite{friedman2008glasso}. Indeed, \cite{barber2020rob} proved that, under certain conditions, the graphical lasso can be used to preserve the FDR guarantees of knockoff filters. For both our model and the classical knockoff filter, we fix the FDR threshold at $q = 0.1$.  The knockoffs were generated using the R package \texttt{knockoffs}, and the covariance matrix was estimated using the \texttt{glasso} package. 
In addition to the model-X knockoff filter, our comparison includes other state-of-art methods for variable selection. In particular, we include the Lasso regression \citep{tibshirani1996lasso} and a Bayesian linear regression model with discrete spike-and-slab prior and a Beta hyperprior on the inclusion probability, which we implemented ourselves. We also considered the joint edge and regression estimation model proposed by \cite{peterson2016joint}, which we implemented ourselves as a version of our proposed model {latent knockoffs layer  and a Bayesian variable selection model with non-local spike and slab proposed by \cite{johnson2012bayesian} and implemented in the R package \texttt{mombf}}. The regularizing parameter for lasso regression was determined {as the largest value such that the cross-validation error is within one standard error of the minimum}, using the R package \texttt{glmnet}. We note that this method does not employ any FDR correction.  Variable selection for the Bayesian methods was achieved  by thresholding the marginal posterior probabilities of inclusion (PPIs) at $0.5$, following the median model approach \citep{barbieri2004optimal}. We also considered the BFDR approach proposed by \cite{newton2004bfdr}. 

To compare performance, in Table~\ref{tab:sim} we report averages and standard deviations across $100$ replicates of several evaluation metrics. In particular, we consider FDR, True Positive Rate (TPR), Matthews Correlation Coefficient (MCC), and F1-score. Let TP be the number of true positives, TN the number of true negatives, FP the number of false positives, and FN the number of false negatives. We can then define the performance metrics used as
\begin{align*}
    & FDR = \frac{FP}{FP + TP}, \qquad \text{TPR}=\frac{TP}{TP + FN}, \\
    & MCC = \frac{TP\times TN - FP \times FN}{\sqrt{(TP+FP)(TP+FN)(TN+FP)(TN+FN)}}, \\
    & F1 = \frac{2\times TP}{2\times TP + FP + FN}.   
\end{align*}
Results in Table~\ref{tab:sim} indicate that the proposed Bayesian method succeeds in keeping the FDR below the given threshold in every scenario, while achieving strong performances in TPR compared to the other methods, as well as MCC and F1 scores. It is worth noting that the classical knockoff filter exhibits low TPR in Scenarios $1$ and $2$, where the magnitude of the true coefficients is rather small. 
%In Scenario $3$, where the signal strength is stronger, the small sample size affects the TPR of the classical knockoff filter. 
In other settings, not reported here, where the signal was strong and the sample size larger, we noticed improved performances for all methods, including classical knockoffs. Further, the values of the standard deviations for the TPR of the knockoff filter in the first two scenarios are higher than those of the other methods. This happens because, across the $100$ simulated data sets, the knockoff filter TPR is often zero and sometimes quite high. 
%We can see that this is not the case when looking at the results for Scenario $3$. With larger true values of the regression coefficient classical knockoff filter has sufficient TPR to identify the true predictors. Therefore, the variability of TPR  decreases. 
Finally, we note that classical knockoff has a slightly higher FDR than the threshold $0.1$ when the covariates are not independent and the covariance matrix is estimated using graphical lasso. As shown in \cite{barber2020rob},  the classical knockoff filter does not control FDR exactly below $q$ when the covariance matrix used to build the knockoffs is not the true one but an estimate. In our case, the graphical lasso tends to produce a sparser than needed graph. This behavior also supports our choice of setting the parameter of the priors in the proposed method to obtain a generally dense graph to be used to generate the knockoffs. Results for the third scenario confirm that the proposed Bayesian Knockoff filter is also able to outperform the state-of-art methods in settings with larger dimensionality. As a matter of fact, it successfully controls FDR under $0.1$, unlike the lasso regression. Moreover, the TPR is on average comparable to the one of the classical knockoff filter, but it is more stable as it has a much lower standard deviation.

\begin{table}
\centering
\caption{{\bf Simulation study:} FDR, TPR, MCC and F1-score, for the scenarios described in Section~\ref{sec:sim_gen}. Compared methods are the proposed model (BayesKnock), classical Knockoff filter with exact (Knockoff (Exact)) and estimated (Knockoff (Glasso)) covariance matrix, Lasso, spike-and-slab regression (Spike-Slab), the joint model proposed by \cite{peterson2016joint} (Joint) and {the non local prior for variable selection proposed by \cite{johnson2012bayesian}}. Values reported are averages over $100$ replicated data sets, with standard deviations in parentheses.}
\begin{tabular}{@{}ll|ll|ll|ll@{}}
\toprule
                                           &                   & \multicolumn{2}{l}{Scenario 1} & \multicolumn{2}{l}{Scenario 2}  & \multicolumn{2}{l}{Scenario 3} \\ \midrule
\multirow{6}{*}{FDR}   & BayesKnock               &  0.007    & (0.03)      &  0.013      & (0.05)   & {\bf 0.032}  & (0.05)      \\
                       & Knockoff (Exact)        & 0.035      & (0.11)      & 0.025      & (0.09)        & 0.055  & (0.08)      \\
                       & Knockoff (Glasso)       & 0.070      & (0.16)      & 0.114      & (0.19)    & 0.079  & (0.08)      \\
                       & Lasso   & 0.146       & (0.15)      & 0.216      & (0.15)    & 0.101   & (0.12)       \\
                       & Spike-Slab              & {\bf 0.000}      & (0.00)      & {\bf 0.003}      & (0.02)     & 0.716  & (0.04)      \\
                       & Joint & 0.029 & (0.06) & 0.043 & (0.07) & 0.177  & (0.11)  \\
                       & Non-local prior        & {\bf 0.001}  & (0.01) & {\bf 0.003} & (0.02) & {\bf 0.000}  & (0.00) \\
                       
\hline
\multirow{6}{*}{TPR}   & BayesKnock              & {\bf 1.000}      & (0.00)      & 0.977      & (0.07)      & 0.560  & (0.10)      \\
                       & Knockoff (Exact)        & 0.088      & (0.28)      & 0.068      & (0.25)     & 0.592  & (0.36)      \\
                       & Knockoff (Glasso)       & 0.162      & (0.37)      & 0.277      & (0.45)    & 0.595  & (0.39)      \\
                       & Lasso   & {\bf 1.000}      & (0.00)      &   {\bf 1.000}    &  (0.00)    & {\bf 0.795}  &  (0.08)   \\
                       & Spike-Slab              & 0.999      & (0.01)      & 0.979      & (0.06)     & {\bf 0.820}  & (0.06)      \\ 
                       & Joint & {\bf 1.000} & (0.00) & {\bf 1.000} & (0.00)  & 0.664 & (0.08)  \\
                       & Non-local prior        & {\bf 1.000} & (0.00) & 0.995  & (0.03) & 0.215  &  (0.05)\\
\hline
\multirow{6}{*}{MCC}   & BayesKnock              & 0.995      & (0.02)      & 0.973      & (0.06)    & {\bf 0.714}  & (0.07)      \\
                       & Knockoff (Exact)        & 0.065      & (0.21)      & 0.059      & (0.20)       & 0.627  & (0.37)      \\
                       & Knockoff (Glasso)       & 0.109      & (0.25)      & 0.188      & (0.31)      & 0.612  & (0.41)      \\
                       & Lasso   & 0.898       & (0.11)      & 0.847      & (0.11)     & {\bf 0.825}   & (0.08)   \\
                       & Spike-Slab              & {\bf 0.999}      & (0.01)      & {\bf 0.986}      & (0.05)       & 0.389 & (0.05)      \\
                       & Joint & 0.981 & (0.04) & 0.972 & (0.05)  & 0.711  & (0.08)  \\
                       & Non-local prior        & {\bf 0.999} & (0.01) & {\bf 0.995}  &  (0.02) &  0.442 &  (0.05)\\
\hline
\multirow{6}{*}{F1}    & BayesKnock        & 0.996  & (0.02) & 0.977 & (0.06)  & 0.704  & (0.08)\\
                       & Knockoff (Exact)  & 0.069 & (0.22) & 0.062 & (0.21)    & 0.631  & (0.37)      \\
                       & Knockoff (Glasso)  & 0.117 & (0.27) & 0.201 & (0.33)     & 0.629  & (0.38)      \\
                       & Lasso   & 0.914      & (0.09)      & 0.871      & (0.10)     & {\bf 0.836}   &    ( 0.07) \\
                       & Spike-Slab        & {\bf 0.999} & (0.01) & {\bf 0.988} & (0.04)  & 0.420  & (0.04)\\ 
                       & Joint & 0.985 & (0.03) &  0.977 & (0.04) & {\bf 0.729} & (0.07)  \\
                       & Non-local prior        & {\bf 0.999}  & (0.01)  & {\bf  0.996 } & (0.02) &0.352   & (0.07) \\\bottomrule
\end{tabular}
\label{tab:sim}
\end{table}

As for the comparison with the state-of-art methods for variable selection, we can see that Lasso has a high TPR in every scenario { but fails to achieve any explicit control over the FDR, even in the simplest scenario.} Compared with Bayesian variable selection methods via spike-and-slab priors, our model offers comparable results in the first two scenarios, where the covariates are independent, with almost perfect results. In the third scenario, with dependent covariates, classical spike-and-slab loses the control of FDR, unlike our method,  even though it manages to find more relevant predictors than our method, while non-local spike-and-slab still controls the FDR but at the cost of a much lower TPR than our method.
In Table \ref{tab:sim} we do not show results for the BFDR approach proposed by \cite{newton2004bfdr}. Indeed, BFDR selected exactly the same variables as spike-and-slab in the first two scenarios. In the third scenario, however, the PPIs were hovering around $0.5$ and the BFDR did not select any variable. As pointed out in the introduction section, Bayes FDR selection procedures rely on the estimates of the marginal PPIs, as obtained from the MCMC output, while our procedure employs an upper bound on the probabilities of non-inclusion. Therefore, our model appears to achieve good operating characteristics and to control the FDR even when the estimates of the marginal posterior probability of inclusion fail to select the true relevant variables. 
This is evident when comparing our method with the model proposed by \cite{peterson2016joint} (Joint). In scenarios 1 and 2 the two methods have the same performances, while in scenario 3, even though MCC and F1 have similar values, the addition of the knockoffs allows to keep the FDR under the desired level, while the joint method has higher FDR and therefore not an explicit control over it.

We note that, with our proposed method, posterior inference results not only in variable selection but also in learning the underlying graph structure among the covariates via the inference on the precision matrix $\bm\Omega$. For example, the median graph can be obtained by thresholding the marginal posterior probability of inclusion (PPI) of edge inclusion at 0.5. This approach of using a threshold for PPIs is commonly employed in graph structural learning tasks as the space of possible graphs is too big to be explored entirely, making it impossible to select the most frequently encountered graph. When looking at the Frobenius norm of the difference between the posterior mean of $\bm \Omega$ and the true precision matrix,
\[
    \Vert \bm \Omega - \bm{\widehat{\Omega}}\Vert_F = \sqrt{\sum_{i, j} \left(\omega_{ij} - \hat\omega_{ij}\right)^2},
\]
our model achieved very solid performances for graph selection across the scenarios. For example, the average Frobenius norm was $1.404$, $1.415$ and $9.425$ and the F1-score was $0.869$,  $0.865$ and $0.99$ in Scenarios $1$, $2$ and $3$, respectively.  Note that the Frobenius norm is much higher in Scenario $3$ due to the different magnitude in the correlation between variables. However, as we can see from the F1-score, the graph recovery still works.

%\subsection{Simulation of a survival outcome from an accelerated failure time model}\label{sec:aft}

We also show results on a simulation from an AFT model for survival outcome, with censored observations.
%To use the proposed Bayesian knockoff filter, we adapt the stochastic search MCMC procedure of \cite{sha2006bayesian}, that makes use of data augmentation techniques \citep{tanner1987calculation}. Let $T_i$ represent the time-to-event for the $i$-th subject and $c_i$ the censoring time. AFT models assume a multiplicative effect on the survival times of the general form $\log (T_i) = X_i \beta + \epsilon_i$, where the errors $\epsilon_i$ may have several distributions. Here, we assume the errors to be i.i.d. Gaussian variables.  To allow for censored observations, we introduce a latent variable $y_i$ such that
%\[
%\begin{cases}
%    y_i = \log(T^*_i) & \text{if } \kappa_i = 1 \\
%    y_i > \log(T^*_i) & \text{if } \kappa_i = 0 \\
%\end{cases}.
%\]
%where $T^*_i$ is defined as $\min(T_i, c_i)$ and $\kappa_i = \mathbbm{1}_{T_i \le c_i}$ is the censoring indicator. 
%Here, we assume that the latent variable $y_i$ given the covariates $X$ follows the model described in Equation~\eqref{eq:pY} with $p(Y \mid X, U)$ as in Equation \eqref{eq:work_U}, with the same prior specification. 
We generate the covariates following the scheme employed in Scenario~$3$ above, and then generate the log of the survival times from the linear AFT model. We allow for $25$ \% of the subjects to be censored. Table \ref{tab:aft} shows performances of our proposed method, compared with the spike-and-slab method for AFT models of \cite{sha2004bayesian}, the same AFT model without \bcolch the latent knockoff layer \ech, as also implemented by \cite{peterson2016joint} { and the non-local spike-and-slab prior as proposed by \cite{johnson2012bayesian}}. These results are consistent with those for Scenario 3 in the previous simulation and also show that the proposed procedure suffers the least from the presence of censored data.

\begin{table}
\centering
\caption{{\bf Simulation study:} FDR, TPR, MCC and F1-score, for the AFT models. Compared methods are the proposed model (BayesKnock), spike-and-slab regression (Spike-Slab), the joint model proposed by \cite{peterson2016joint} (Joint) and { the non local prior for variable selection proposed by \cite{johnson2012bayesian}.} Values reported are averages over $100$ replicated data sets, with standard deviations in parentheses.}
\begin{tabular}{l|llllllll}
\toprule
           & \multicolumn{2}{l}{FDR} & \multicolumn{2}{l}{TPR} & \multicolumn{2}{l}{MCC} & \multicolumn{2}{l}{F1} \\
           \midrule
BayesKnock & 0.035      & (0.03)     & 0.526      & (0.11)     & {\bf 0.713}      & (0.08)     & 0.701     & (0.10)     \\
Spike-Slab & 0.642      & (0.04)     & {\bf 0.742}      & (0.01)     & 0.239      & (0.08)     & 0.311     & (0.08)     \\
Joint      & 0.192      & (0.13)     & 0.592      & (0.04)     & 0.701      & (0.09)     & {\bf 0.712}      & (0.10) \\ 
Non-local prior & {\bf 0.000} & (0.00) & 0.203 & (0.07) & 0.409 & (0.08) & 0.344 & (0.12) \\
\bottomrule
\end{tabular}
\label{tab:aft}
\end{table}

\subsection{Sensitivity analysis}
\label{sec:sens}
Table~\ref{tab:sensitivity} shows the results of the sensitivity analysis. We started from the same setting adopted in the simulation study for Scenario 2 and then varied one hyperparameter at a time. We found that posterior inference is robust to the choice of the hyperparameters, except for $v_0$ in the prior on the precision matrix $\bm \Omega$ in Equation~\eqref{priorOmega}. Both \cite{wang2015graph} and  \cite{peterson2016joint} report the same sensitivity. We notice, however, that the choice of $v_0$ does not affect the selection performances of our method. As noted in \cite{barber2020rob}, knockoffs can still be used with estimated distributions, but the control on FDR is affected by the Kullback-Leiber distance between the estimate and the true distribution. As a matter of fact, in Gaussian graphical models, imposing too many zeros strongly affects the estimates of the non-zero off-diagonal elements on the precision matrix $\bm \Omega$. For this reason, we recommend setting the hyperparameters $v_0$ and $v_1$ to obtain a slightly denser graph. 

%  \multicolumn{5}{c}{$(h_\beta=1 \quad v_0 = 0.01^2 \quad v_1 = 1  \quad \xi = 0.01 \quad b = 0.5 \quad a_\sigma =2 \quad b_\sigma = 2)$}\\  

\begin{table}
    \centering
    \caption{{\bf Simulation study:} Sensitivity analysis.}
    \label{tab:sensitivity}
    \begin{tabular}{l|cc|cc}
    \toprule
    Hyperparameter & $\quad$FDR$\;$ & $\;$TPR$\quad$ & Frobenius Loss & F1 Score \\
    \hline
    $a=-2.5$ & 0.000 & 1.000 & 1.624 & 0.800\\
    $a=-0.5$ & 0.000 & 1.000 & 1.617 & 0.840\\
    $a=0.5$  & 0.000 & 1.000 & 1.607 & 0.824\\
    $a=2.5$  & 0.142 & 1.000 & 1.662 & 0.800\\
    \hline
    $b=-2.5$ & 0.000 & 0.833 & 1.625 & 0.816\\
    $b=-0.5$ & 0.000 & 0.833 & 1.600 & 0.840\\
    $b=0.5$  & 0.000 & 1.000 & 1.607 & 0.824\\
    $b=2.5$  & 0.333 & 1.000 & 1.645 & 0.816\\
    \hline
    $v_0=0.01^2$ & 0.000 & 1.000 & 1.620 & 0.816\\
    $v_0=0.1^2$  & 0.000 & 1.000 & 2.435 & 0.067\\
    $v_0=0.5^2$  & 0.000 & 1.000 & 4.134 & 0.000\\
    \hline
    $v_1=0.25$   & 0.000 & 1.000 & 1.525 & 0.840\\
    $v_1=1.00$   & 0.000 & 1.000 & 1.645 & 0.816\\
    $v_1=25.00$  & 0.000 & 1.000 & 1.708 & 0.840\\
    $v_1=100.00$ & 0.000 & 1.000 & 1.901 & 0.766\\
    \bottomrule
    \end{tabular}
\end{table}

\section{Application to prostate cancer data} \label{Sec:4_applic}
As an illustrative example, we examine a data set on prostate cancer obtained from \cite{stamey1989prostate}  and widely used as a benchmark data set \citep[see, for instance,][among others]{tibshirani1996lasso, zou2005elnet}. The aim is to investigate the association between Prostate-Specific Antigen (PSA) levels and various clinical measures in a sample of $97$ men, who were preparing to undergo radical prostatectomy. \texttt{PSA} is a protein produced by normal and malignant prostate cells and is useful as a preoperative marker as prostate cancer releases PSA into the bloodstream. Additionally, eight clinical measures are included: the logarithm of the cancer volume (in cm$^3$) (\texttt{lcavol}), the logarithm of the prostate weight (in g) (\texttt{lweight}), patient age (\texttt{age}), The logarithm of the amount of benign prostatic hyperplasia (in cm$^2$) (\texttt{lbph}), the logarithm of capsular penetration (in cm) (\texttt{lcp}), seminal vesicle invasion (\texttt{svi}), the current Gleason score (\texttt{gleason}) and the percentage of Gleason scores four or five out of five (\texttt{pgg45}). The logarithm of the PSA (ng/ml) is the response variable. The presence of seminal vesicle invasion is a binary variable (1 = yes, 0 = no) and \texttt{gleason} is a discrete numerical variable with four values. The Gleason score relates to the current grade of prostate cancer, while the predictor \texttt{pgg45} provides information about the patient's history of the Gleason scores, and it is strongly linked to the current final score  Gleason score. 

For our analysis, we considered the subset of $73$ subjects without seminal vesicle invasion and with no severe cancer (Gleason score less than $9$). 
To ensure the appropriate statistical assumptions, we applied the nonparanormal transformation \citep{liu2009} to the covariates. This transformation provides a Gaussian marginal distribution of each covariate and assumes a Gaussian copula on their joint distribution. We tested the joint Gaussian assumption after this transformation using the multivariate Shapiro test \citep{villasenor2009generalization}, obtaining a $p$-value of $0.02$. In our application, we used the same hyperparameter settings as in the simulation study, and ran an MCMC chain with $8,000$ iteration for burn-in followed by $8,000$ iterations for inference. The average $z$-scores from the Geweke's test for $\{(\beta_j, \widetilde{\beta}_j)\}_{j=1}^p$ was $-0.22$.

\begin{figure}[h]
    \centering
    %left, bottom, right, top
    \includegraphics[trim = 0cm 1cm 0cm 2cm, clip, width = 0.7\textwidth]{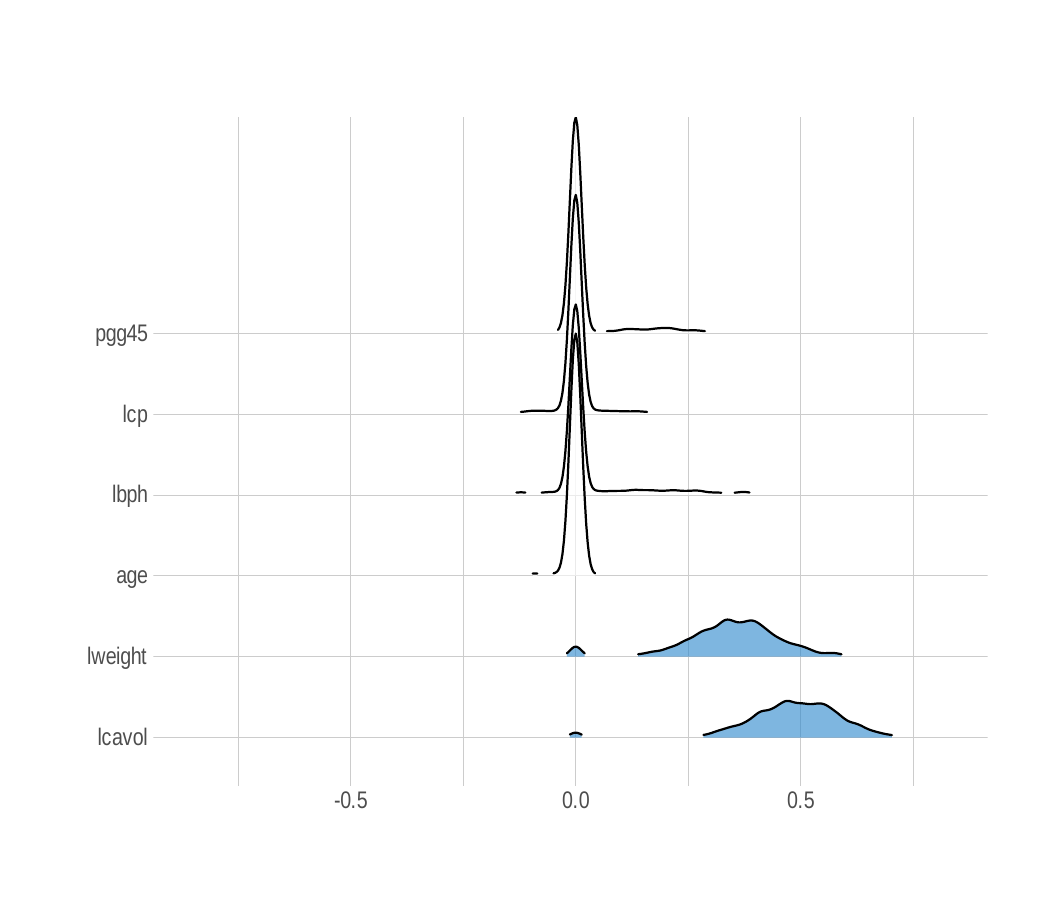}
    \caption{{\bf Prostate cancer data}: Posterior distribution of $W_j$, $j=1, \dots, 6$. Those highlighted in black are selected according to Figure~\ref{fig:selection}}
    \label{fig:post}
\end{figure}

\begin{figure}[h]
    \centering
    \includegraphics[trim = 0cm 1cm 0cm 1cm, clip,width = 0.9\textwidth]{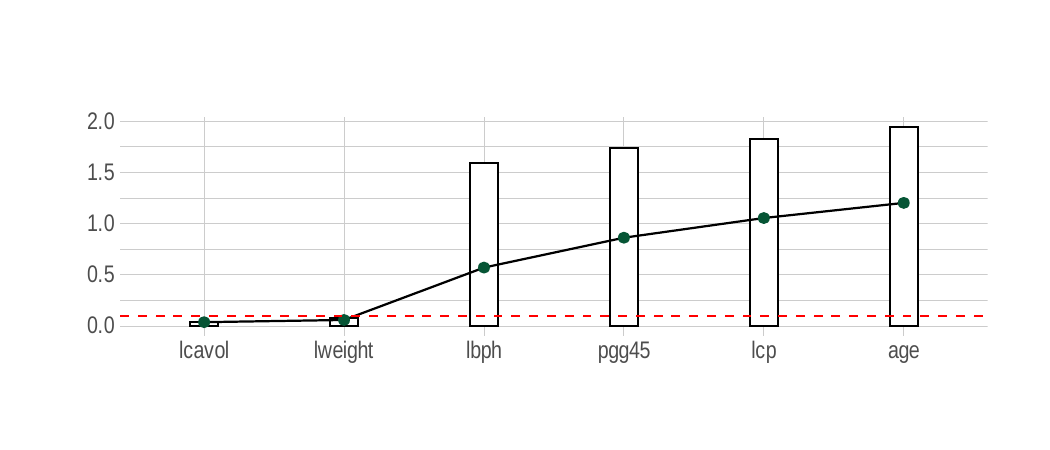}
    \caption{{\bf Prostate cancer data}: Variable selection. Bars represent estimates $2\, \widehat{\mathbb{P}}[W_j \le 0\mid \bm D]$, $j = 1, \ldots 6$, in increasing order and points represent the estimated $\text{BFDR}(\mathcal{S})$ with $\mathcal{S}$ the set of all previous indexes. The dashed red line is the chosen threshold $q=0.1$.}
    \label{fig:selection}
\end{figure}

Figure~\ref{fig:post} shows the posterior distribution of the feature statistics $W_j$, $j=1, \ldots, p$ for each covariate in the model, with the selected variables shown in black. Notably, the posterior distribution of the non-selected variables is symmetric around zero, while the selected variables exhibit a positive-centered distribution. Additionally, Figure~\ref{fig:selection} reports the estimated values of $2\, \widehat{\mathbb{P}}[W_j \le 0\mid \bm D]$, $j = 1, \ldots 6$, in increasing order, together with the corresponding estimated $\text{BFDR}(\mathcal{S})$. Therefore, setting $q=0.1$, the covariates to be selected are the logarithm of cancer volume and the logarithm of prostate weight. 

We also applied the state-of-art procedures used for comparison in the simulation study. Table~\ref{tab:comp} summarizes the results. We can see that  classical knockoff filter does not find any association between covariates and PSA production level. {Classical and non-local} spike-and-slab regressions select the same covariates as the proposed method, and the joint estimation model selects one additional variable. Lasso, as we saw in the simulation study, selects many variables, including the two selected by our method. We acknowledge that the Bayesian procedures are computationally more expensive than the classical methods. For example, running the MCMC for our proposed method on this application took 35 minutes on a regular laptop equipped with Intel(R) Core(TM) i7-9750H CPU @ 2.60GHz. However, relying on a posterior distribution of the latent knockoff variables increases stability of the selection, as our simulations have shown.
%{\color{green} On the same laptop the spike-and-slab regression took 1 minute and the joint method 5 minutes. Note that in this application the impact of knockoffs on the computational cost is quite high due to the low number of covariates. As a matter of fact, when $p$ increases the cost of sampling the graph becomes much higher so that the joint method and the proposed one has similar computational cost.}
%{\color{green} La spike and slab ha praticamente zero di computing time perché integrano via i beta, non so se conviene metterlo.}

\begin{table}
\caption{{\bf Prostate cancer data}: Selection of variables using different variable selection methods.}
\label{tab:comp}
\centering
\begin{tabular}{lcccccc}
\toprule
Variable & BayesKnock                & Knockoff Filter & Lasso                     & Spike-and-slab & Joint & Non-local\\
\midrule 
lcavol   & \checkmark & -               & \checkmark & \checkmark             & \checkmark & \checkmark\\
lweight  & \checkmark & -               & \checkmark & \checkmark              & \checkmark & \checkmark\\
age      & -                         & -               & -                         & -          & -    & -\\
lbph     & -                         & -               & \checkmark & -            & \checkmark  & -\\
lcp      & -                         & -               & -                         & -         & -  &-    \\
pgg45    & -                         & -               & \checkmark & - & - &-\\ \bottomrule           
\end{tabular}
\end{table}

\section{Concluding remarks}
\label{Sec:5_concludo}

In this article, we have proposed a fully Bayesian generalization of  the model-X knockoff filter of \cite{candes2018gold} that allows for variable selection while controlling the BFDR at an arbitrary level. 
Our framework \bcolch introduces a latent knockoff layer through a parameter-expanded representation of the response model, where knockoffs act as matched negative controls for feature scoring. \ech 
Additionally, we have imposed a sparse conditional independence structure on the covariates and used it in the prior on the \bcolch latent \ech knockoffs in a way that satisfies their pairwise conditional independence property. We have also limited the computational cost by adopting a modified version of the spike-and-slab prior that avoids the increase of the model dimension. 
Our model performs variable selection using an upper bound on the posterior probability of non-inclusion. \bcolch We have shown that the latent knockoff variables generated by our construction satisfy the defining properties of valid model-X knockoffs a priori and that the resulting variable-selection procedure controls the BFDR at an arbitrary level in finite samples when the covariate distribution is fully known. When the precision matrix is estimated under the proposed graphical model, the procedure still retains an asymptotic FDR guarantee. \ech

%We have shown how our model construction leads to valid model-X knockoffs and demonstrated that the proposed approach is sufficient for controlling the BFDR at an arbitrary level, in finite samples, if the distribution of the covariates is fully known, and asymptotically if estimated as in our model. 

In applications, we have shown that our proposal increases the stability of the selection, since it relies on \bcolch 
posterior averaging over the latent knockoff layer rather than on a single realized knockoff sample. \ech Results from our simulation study have shown how in simple settings (like Scenarios 1 and 2) our Bayesian knockoff filter approach has similar performance to state-of-art Bayesian variable selection techniques, such as spike-and-slab priors, while outperforming classical variable selection methods, including classical knockoffs filter. Moreover, under more complex dependence structure of the covariates (as in Scenario 3), our proposal is able to control the FDR under any desired threshold, unlike spike-and-slab variable selection methods, that suffer from poor estimates of the PPIs. It also performs as well as classical knockoff filter with covariance structure known. This is relevant since in practice the precision matrix is never known. 
%As an illustrative example, we have applied the model to a benchmark data set on prostate cancer collected from the study of \cite{stamey1989prostate}, and found more relevant variables than the classical knockoff filter.

Future work could extend this research in several directions. First, going beyond the standard Gaussian linear regression setting, we have also discussed how similar Bayesian knockoff filters can be designed for linear settings that incorporate discrete-type responses, such as AFT models for survival outcomes and probit models for multinomial responses. {More general extensions of the proposed Bayesian knockoff filter formulation might be possible in modeling frameworks that employ latent variables formulations. }
%{\color{teal}For example, logit and negative binomial responses could be accommodated via the use of Polya-Gamma data augmentation schemes \citep{polson2013bayesian, zhou2012lognormal}. }}
Another interesting direction would be to extend the proposed knockoffs framework to the case of non-Gaussian covariates.  The generation of knockoffs for non Gaussian random variables is still an open research problem. This case presents the main challenge of not having a known distribution that respects pairwise exchangeability \citep{bates2021metr}. A promising direction is represented by the approach proposed by \cite{dreassi2024generating}, which makes use of conditional independence to build the knockoffs variables.
%Furthermore, the extension to non-Gaussian covariates of our current proposal to sample knockoffs from a graphical model is an interesting and computationally challenging future avenue.
Finally, the variable selection prior distribution we have used on the coefficients presents some computational challenges due to the need to use a Metropolis-within-Gibbs algorithm. %Since the variable selection does not directly rely on the Posterior Probability of Inclusion (PPI) provided by this type of prior, 
%As an alternative, a shrinkage prior such as the horseshoe prior \citep{Carvalho2010}, could be used to reduce computational costs. However, this reduced computational cost comes at the expense of greater difficulty in utilizing the information from the estimated graph on the covariates. 
\bcolch Since the largest computational burden comes from updating the latent knockoff augmentation at each MCMC iteration, \ech a possible solution would be to use variational inference \citep{blei2017variational}. However,  the restrictive assumptions to obtain valid knockoffs require careful design of the approximate variational distributions. In particular, the commonly used mean field approximation cannot be used since it breaks the pairwise exchangeability assumption of the knockoffs and the covariates. 
%{\color{green} A promising knockoff generation setting to include non-Gaussian covariates is the one proposed by \cite{berti2023new}. Authors, indeed, found a way to generate knockoffs for any random variable as long as it is possible to find some other random variable such that the covariates are conditional independent given the new variable. Thus, it seems natural to include this idea in a Bayesian settings where the parameters of the covariates distribution are treated as random variables.}

\section*{Software Availability}
Code reproducing the results of this paper will be available on GitHub upon acceptance of the paper. 
 
\section*{Acknowledgments}
We thank the Editor, Associate Editor, and anonymous referees for their thoughtful and constructive feedback, which greatly improved the quality of this manuscript.\\

\noindent 
This work was performed as part of the first author's PhD thesis, and while the author was visiting scholar at Rice University. The first and second authors were partially supported by the Italian Ministry of University and Research (MUR), Department of Excellence project 2023-2027 ReDS 'Rethinking Data Science' - Department of Statistics, Computer Science, Applications - University of Florence, and the MUR-PRIN grant 2022 SMNNKY, CUP B53D23009470006 European Union - NextGenerationEU, Mission 4 Component, 2. The second author was also partially supported by the European Union - NextGenerationEU - National Recovery and Resilience Plan, Mission 4 Component 2 - Investment 1.5 - THE - Tuscany Health Ecosystem - ECS00000017 - CUP B83C22003920001.

\bibliographystyle{apalike}
\bibliography{biblio.bib}       % Bibliography file (usually '*.bib')

\end{document}